\documentclass[footinbib,aps,prl,reprint,superscriptaddress]{revtex4-1}

\usepackage{amsmath,amssymb,braket,upgreek,bm,verbatim}
\usepackage{graphicx,tabularx} 
\usepackage{xcolor}
\definecolor{darkblue}{rgb}{0,0,0.5}
\usepackage{hyperref}
\hypersetup{
colorlinks=true,
linkcolor=black,
filecolor=blue,
citecolor=darkblue,  
urlcolor=black,
}

\usepackage{multirow}
\usepackage{booktabs}
\usepackage{xcolor}

\usepackage[absolute]{textpos}
\usepackage{xcolor}
\usepackage{multirow}

\usepackage{amsmath,amssymb,bbm,bm}
\usepackage{mathtools}
\usepackage{braket}
\usepackage{amsmath}
\usepackage{relsize}
\usepackage{comment}
\usepackage{csquotes}

\newcommand{\appropto}{\mathrel{\vcenter{
  \offinterlineskip\halign{\hfil$##$\cr
    \propto\cr\noalign{\kern2pt}\sim\cr\noalign{\kern-2pt}}}}}

\begin{document}

\title{Complete frequency-bin Bell basis synthesizer}


\author{Suparna Seshadri}
\affiliation{Elmore Family School of Electrical and Computer Engineering and Purdue Quantum Science and Engineering Institute, Purdue University, West Lafayette, Indiana 47907, USA}
\author{Hsuan-Hao Lu}
\affiliation{Quantum Information Science Section, Oak Ridge National Laboratory, Oak Ridge, Tennessee 37831, USA}
\author{Daniel E. Leaird}
\affiliation{Elmore Family School of Electrical and Computer Engineering and Purdue Quantum Science and Engineering Institute, Purdue University, West Lafayette, Indiana 47907, USA}
\author{Andrew M. Weiner}
\affiliation{Elmore Family School of Electrical and Computer Engineering and Purdue Quantum Science and Engineering Institute, Purdue University, West Lafayette, Indiana 47907, USA}
\author{Joseph M. Lukens}
\email{lukensjm@ornl.gov}
\affiliation{Quantum Information Science Section, Oak Ridge National Laboratory, Oak Ridge, Tennessee 37831, USA}







\date{\today}

\begin{abstract}
We report the experimental generation of all four frequency-bin Bell states in a single versatile setup via successive pumping of spontaneous parametric downconversion with single and dual spectral lines. Our scheme utilizes intensity modulation to control the pump configuration and offers turn-key generation of any desired Bell state using only off-the-shelf telecommunication equipment. We employ Bayesian inference to reconstruct the density matrices of the generated Bell states, finding fidelities $\geq$97\% for all cases. Additionally, we demonstrate the sensitivity of the frequency-bin Bell states to common-mode and differential-mode temporal delays traversed by the photons comprising the state---presenting the potential for either enhanced resolution or nonlocal sensing enabled by our complete Bell basis synthesizer. 

\end{abstract}

\begin{textblock}{13.3}(1.4,15)
\noindent\fontsize{7}{7}\selectfont \textcolor{black!30}{This manuscript has been co-authored by UT-Battelle, LLC, under contract DE-AC05-00OR22725 with the US Department of Energy (DOE). The US government retains and the publisher, by accepting the article for publication, acknowledges that the US government retains a nonexclusive, paid-up, irrevocable, worldwide license to publish or reproduce the published form of this manuscript, or allow others to do so, for US government purposes. DOE will provide public access to these results of federally sponsored research in accordance with the DOE Public Access Plan (http://energy.gov/downloads/doe-public-access-plan).}
\end{textblock}

\maketitle

\textit{Introduction.---}Bell states are vital resources both for fundamental investigations of quantum entanglement and for realizing practical goals in quantum information
processing and metrology. Generation and measurement of Bell states appear in a plethora of quantum communication protocols spanning dense coding~\cite{mattle1996dense}, teleportation~\cite{bennett1993teleporting}, entanglement-based cryptography~\cite{shukla2014protocols, shi2013multi, tittel2000quantum}, and entanglement swapping~\cite{pan1998experimental,goebel2008multistage}. 
Production of a complete set of Bell states has been actively studied in various photonic degrees of freedom including polarization~\cite{kwiat1995new,mattle1996dense}, orbital angular momentum~\cite{agnew2013generation}, discrete time-bins~\cite{Lo2020}, pulsed time-frequency modes~\cite{brendel1999pulsed} %
and path encodings~\cite{shadbolt2012generating,silverstone2014chip,li2020femtosecond}. Recently, interest in frequency-bin encoding has grown due to particular advantages such as simple multiplexing capabilities and compatibility with both on-chip integration and optical fiber networks~\cite{kues2019quantum, imany201850, lu2018electro, zhang2021chip}. In frequency bins, the \emph{negative} correlations associated with the $\ket{\Psi^{\pm}}\propto\ket{01}\pm\ket{10}$ Bell states (under the convention where the logical $\ket{1}$ has higher frequency than logical $\ket{0}$ for each photon) are automatically realized through energy conservation in a nonlinear parametric process driven by a continuous-wave (CW) monochromatic pump (or pulsed pump with bandwidth 
less than the frequency-bin separation of interest). However, the generation of \emph{positively} frequency-correlated $\ket {\Phi^{\pm}}\propto\ket{00}\pm\ket{11}$ states is inherently more challenging. While the $\ket{\Psi^{\pm}}$ states can be deterministically transformed to $\ket{\Phi^{\pm}}$ states using a quantum frequency processor (QFP)~\cite{lu2018quantum}, such transformations require multiple active elements after photon generation, increasing complexity and insertion losses.

In this Letter we synthesize all four frequency-bin Bell states in a single setup by successively driving spontaneous parametric down-conversion (SPDC) with single and dual spectral-line pumps. Applying a programmable spectral filter, we then carve out the desired Bell state from the broadband spectrum. We use Bayesian estimation to reconstruct the density matrices of the generated Bell states from mutually unbiased basis measurements and find fidelities $\geq$97\%.
The presented scheme together with the recent demonstration of a frequency-bin Bell state analyzer~\cite{lingaraju2022bell} and arbitrary control of frequency-bin qubits~\cite{lu2020fully} lays the groundwork for several entanglement-based quantum networking protocols. 
Our work also opens up avenues in quantum metrology. We present a proof-of-concept demonstration by probing the opposite impacts of common-mode and differential-mode phases on the positively and negatively correlated Bell states. The results indicate two-photon advantage in the sensitivity of measuring common-mode delays using $\ket{\Phi^{\pm}}$ states which, together with the nonlocal sensing capability of $\ket{\Psi^{\pm}}$ states for differential-mode delays, can be exploited to sense the link latency or perform positioning and clock synchronization~\cite{giovannetti2002positioning, giovannetti2001quantum} in an entanglement-distribution network.

\textit{Background.---}Consider two frequency-bin qubits defined on a comb-like grid of narrowband spectral modes spaced by multiples of $\Delta\omega$. In this context, ``narrowband'' implies that the individual bin widths are smaller than all other characteristic frequency scales in the problem---e.g., variations in the phase-matching function, pulse shaper filter widths, or dispersion. An arbitrary pure two-photon state for an idler $I$ and signal $S$ qubit can then be expressed as
\begin{equation}
\label{eq:arb}    
\ket{\psi} = c_{00}\ket{I_0S_0} + c_{01}\ket{I_0S_1} + c_{10}\ket{I_1S_0}  + c_{11}\ket{I_1S_1},
\end{equation}
where $I_n$ ($S_n$) signifies a single photon populating mode centered at frequency $\omega_{I,n}=\omega_{I,0} + n\Delta\omega$ ($\omega_{S,n}=\omega_{S,0} + n\Delta\omega$). In exploring the feasibility of producing such a state through SPDC, we note immediately that three pump wavelengths will be required to satisfy energy conservation for the four logical states: $\omega_{P,-1}=\omega_{I,0}+\omega_{S,0}$, $\omega_{P,0}=\omega_{I,0}+\omega_{S,1}=\omega_{I,1}+\omega_{S,0}$, and $\omega_{P,1}=\omega_{I,1}+\omega_{S,1}$. Combined with some phase-matching function $\beta$, we therefore can write the basis coefficients $c_{mn}$ in terms of complex pump amplitudes $\alpha_{k}$ corresponding to pump frequency line at $\omega_{P,k}$~\cite{grice1997spectral, grice1997spectral}: $c_{00}=\alpha_{-1}\beta_{00}$, $c_{01}=\alpha_{0}\beta_{01}$, $c_{10}=\alpha_{0}\beta_{10}$, and $c_{11}=\alpha_{1}\beta_{11}$, where $\beta_{mn}\equiv \beta(\omega_{I,m},\omega_{S,n})$ denotes the phase-matching coefficient. 

\begin{figure}[tb!]
  \centering
  \includegraphics[width=\linewidth]{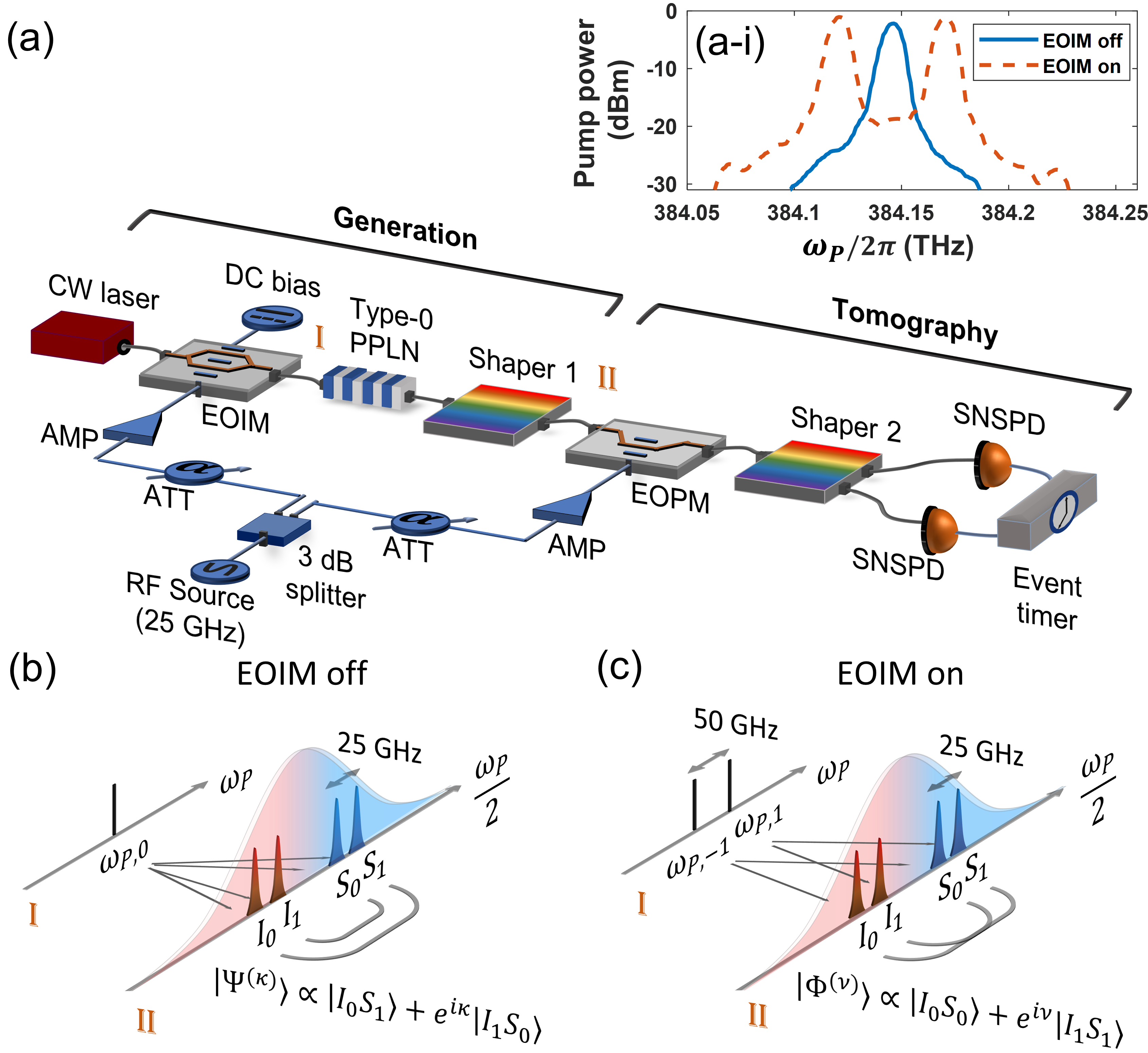}

\caption{(a) Experimental setup. Frequency domain illustration of the scheme for the generation of (b)~$\ket{\Psi^{(\kappa)}}$- and (c)~$\ket{\Phi^{(\nu)}}$-type states. AMP: RF amplifier, ATT: variable RF attenuator, 
EOIM: electro-optic intensity modulator, PPLN: periodically poled lithium niobate, EOPM: electro-optic phase modulator, 
SNSPD: superconducting nanowire single-photon detector. (a-i) Pump spectra measured at 1~GHz resolution.}
\label{setup}
\end{figure}

Accordingly, with full control of three pump lines and the relevant phase-matching conditions, it is in principle possible to produce any two-qubit frequency-bin pure state. Yet despite a large body of research on engineered periodic poling designs for nonlinear crystals~\cite{hum2007quasi}, which in the context of biphoton generation include chirped patterns for increased bandwidth~\cite{Harris2007, Nasr2008, Sensarn2010}, phase-modulated patterns for pump switching~\cite{Odele2015, Odele2017}, and Gaussian patterns to remove spectral entanglement~\cite{Branczyk2011, BenDixon2013, Chen2017}, we are unaware of any method to leverage such engineering for fully arbitrary control over the $\beta_{mn}$ phase-matching factors, as would be required for general two-qubit frequency-bin states. Indeed, if seeking a single physical configuration to produce all states of interest, the condition $|\beta_{00}|\approx|\beta_{01}|\approx|\beta_{10}|\approx|\beta_{11}|$ would actually prove desirable in enabling comparable efficiency for all logical basis states. Under this condition, direct production of an arbitrary state is no longer possible, for the dependence of both $c_{01}$ and $c_{10}$ on $\alpha_{0}$ prevents independent specification of each coefficient~\cite{SFWM}. Nevertheless, all four Bell states can be generated: $\alpha_{\pm1}=0$ and $\alpha_{0}\neq 0$ leads to $|c_{00}|=|c_{11}|=0$ and $|c_{01}|=|c_{10}|\neq 0$, whereas $|\alpha_{1}|=|\alpha_{-1}|\neq0$ and $\alpha_{0}= 0$ yields $|c_{00}|=|c_{11}|\neq0$ and $|c_{01}|=|c_{10}|=0$. By applying phase shifts using a pulse shaper after generation---which would likely be present already for subsequent routing and processing---the specific phases required for $\ket{\Psi^\pm}$ and $\ket{\Phi^\pm}$ can be realized. Importantly, since switching between $\ket{\Psi^{\pm}}$ and $\ket{\Phi^{\pm}}$ is effected by modifying the pump, the generated photons experience no additional loss in the process, in contrast to using an active QFP to convert from $\ket{\Psi^\pm}$-type to $\ket{\Phi^\pm}$-type correlations~\cite{lu2018quantum}. 

\textit{Bell state demonstration.---}
Figure~\ref{setup}(a) depicts our frequency-bin Bell basis synthesizer. A CW laser centered at 780.3~nm ($\omega_{P,0}/2\pi = 384.15$~THz) is launched into an electro-optic intensity modulator (EOIM) driven by a 25~GHz radio-frequency (RF) sinusoidal waveform, set to one of two desired modes of operation: 
in the ``EOIM off'' case, the RF waveform is suppressed and the DC bias point adjusted for maximum transmission, resulting in a single pump line; in the ``EOIM on'' case, an RF waveform with 
a $3.6$~V peak amplitude---approximately 
70\% of the EOIM's half-wave voltage---is applied while the DC bias point is set to the null transmission point, leading to two spectral lines spaced at 50~GHz via carrier suppression.
The output from the EOIM is used to pump a fiber-pigtailed periodically poled lithium niobate (PPLN) ridge waveguide engineered for type-0 phase matching and temperature-tuned to $\sim$56~$^\circ$C for maximum efficiency at the pump frequency $\omega_{P,0}$.  Two 14~GHz-wide frequency bins separated by $\Delta \omega/2\pi = 25$~GHz are carved using a pulse shaper (Shaper 1) at spacings of $\pm$152.5~GHz (for $I_1$ and $S_0$) and $\pm$177.5~GHz (for $I_0$ and $S_1$) on either side of the CW laser's half-frequency ($\frac{1}{2}\omega_{P,0}$). 

\begin{figure}[t]
\centering 
\includegraphics[width=\linewidth]{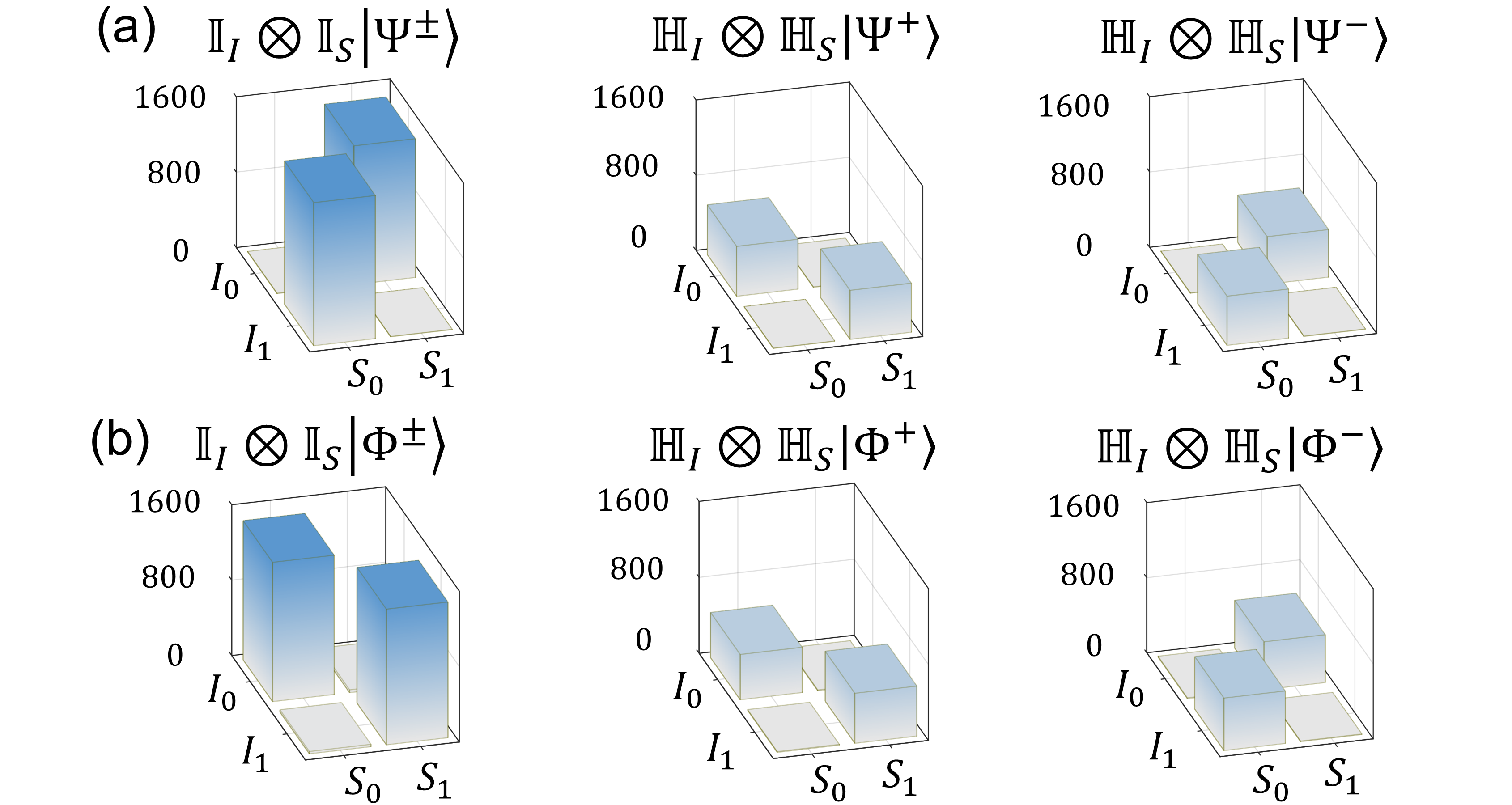}
\caption{Coincidences (integrated over 4~s) between output frequency bins after the gate operations $\mathbb{I}_I\otimes\mathbb{I}_S$ and $\mathbb{H}_I\otimes\mathbb{H}_S $ for  (a)~$\ket {\Psi^{\pm}}$ and (b)~$\ket{\Phi^{\pm}}$ states, respectively.}
\label{JSI}
\end{figure}

The produced state is then characterized by a tomography setup comprising an electro-optic phase modulator (EOPM), a pulse shaper operated as a wavelength-selective switch (Shaper 2), and two superconducting nanowire single-photon detectors (SNSPDs) for coincidence detection. When the EOPM drive signal is off, the measured joint spectral intensity (JSI) corresponds to application of the identity to both signal and idler photons ($\mathbb{I}_I\otimes\mathbb{I}_S$); when the 25~GHz EOPM signal is on (with modulation index $m = 1.435$~rad to ensure equal mixing probability between two adjacent bins), both photons experience a probabilistic Hadamard gate ($\mathbb{H}_I\otimes\mathbb{H}_S$)~\cite{imany2018frequency,lu2020fully} prior to spectrally resolved detection. 

In the first experiment [Fig.~\ref{setup}(b)], 
we couple a single carrier pump ($\omega_{P,0}$) at 6.2~mW input into the PPLN waveguide. 
After spectral filtering, the resulting state is ideally of the form $\ket{\Psi^{(\kappa)}} \propto \ket{I_0S_1} + e^{i\kappa}\ket{I_1S_0}$, 
where the phase $\kappa$ is dependent on the difference in delays experienced by the biphotons prior to the EOPM~\cite{Supplement}. Since both photons traverse identical links with negligible dispersion, $\kappa$ is expected to be zero. 
We verify the same by measuring the coincidences between all pairs of signal and idler frequency bins selected by Shaper 2 while the spectral phase on the bin pair $\ket{I_1S_0}$ is scanned by Shaper 1 followed by parallel Hadamard gates applied by the EOPM. 
The phase $\kappa$ is set to $0$ ($\pi$) using Shaper 1 to obtain the standard Bell state $\ket{\Psi^{+}}\equiv\ket{\Psi^{(0)}}$ ($\ket{\Psi^{-}}\equiv\ket{\Psi^{(\pi)}}$). 

Figure~\ref{JSI}(a) shows the measured coincidences after applying gate operations $\mathbb{I}_I\otimes\mathbb{I}_S$ and $\mathbb{H}_I\otimes\mathbb{H}_S$---corresponding to measurement in the $Z \otimes Z$ and $X\otimes X$ Pauli bases, respectively. In accordance with theory~\cite{Supplement}, the negative frequency correlations revealed in the $\mathbb{I}_I\otimes\mathbb{I}_S$ JSI are reversed after the Hadamards for the $\ket{\Psi^+}$ Bell state, but retained for $\ket{\Psi^-}$. A factor of $\sim$2.8 lower coincidences for the $\mathbb{H}_I\otimes\mathbb{H}_S$ cases result from the probabilistic nature of the single-EOPM Hadamard~\cite{imany2018frequency}, and would not be observed with a full QFP version~\cite{lu2018quantum}. 

\begin{figure}[tb!]
\centering 
\includegraphics[width=\linewidth]{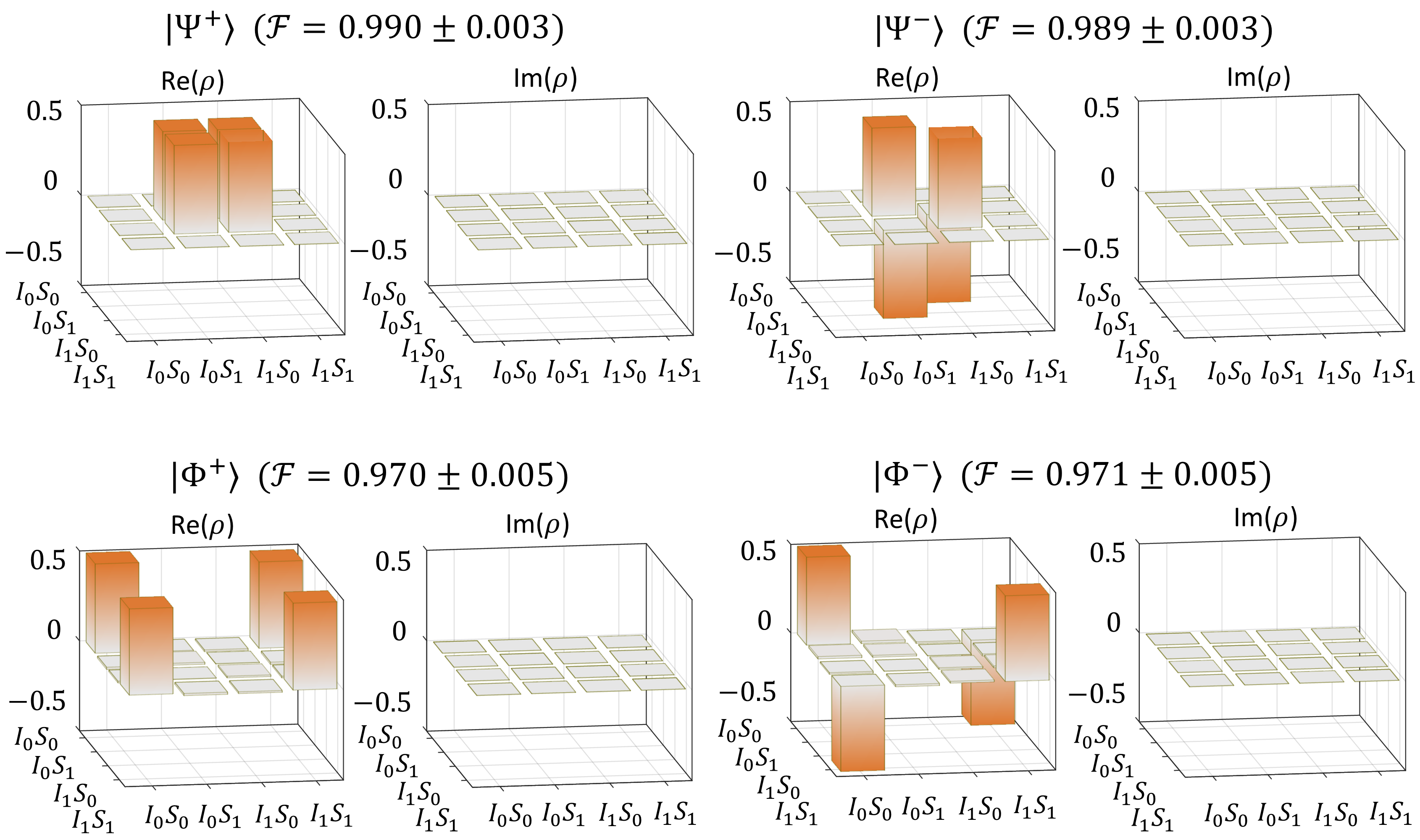}
\caption{ Real and imaginary parts of the Bayesian-mean-estimated density matrices of all four Bell states, computed from the measurements in Fig.~\ref{JSI}.}
\label{Rho}
\end{figure}

In the second experiment [Fig.~\ref{setup}(c)], 
the EOIM eliminates the original pump line at $\omega_{P,0}$ and produces equal first-order sidebands spaced by 50~GHz ($\omega_{P,-1}$ and $\omega_{P,1}$). The power in each sideband after modulation is maintained at $\sim$7~mW in order to achieve coincidence rates similar to the first experiment. Note that this amounts to approximately twice the total pump power as before: in the spontaneous regime, the flux in any given signal-idler bin pair is directly proportional to the pump power at the corresponding sum frequency, so that the two pump lines in the ``EOIM on'' case must \emph{each} match the power of the single line in the ``EOIM off'' case to maintain the rate of Bell state production. 
Experimentally, we observe 17.5~dB extinction of the original pump line [Fig.~\ref{setup}(a-i)], which implies a roughly 50-fold suppression of negatively correlated biphoton contributions relative to the desired positive correlations. Such intensity modulation offers a particularly simple approach for producing the two lines required, although more general pump inputs would be possible with an optical frequency comb as input---e.g., an EOPM followed by a line-by-line pulse shaper in the 780~nm wavelength band~\cite{Monmayrant2004,Willits2012}. Such an arrangement would allow for arbitrary weightings of the input pump lines, but introduce additional complexity that is not required for the specific Bell state cases of interest here.

The succeeding SPDC process generates time-energy entangled biphotons in coherent superpositions of broadband spectral amplitudes centered at half of the pump-sideband frequencies, 
resulting in a two-qubit entangled state ideally of the form $\ket{\Phi^{(\nu)}} \propto \ket{I_0,S_0} + e^{i\nu}\ket{I_1,S_1}$. The phase 
$\nu$ is a fixed common-mode phase that is expected from the RF modulation phase and mean optical delay traversed by the biphotons~\cite{Supplement}. 
We determine the phase $\nu$ in the same fashion as with $\kappa$ but now by scanning the spectral phase imparted on the bin pair $\ket{I_1S_1}$; the measured value of $\nu$ is compensated for and set to $0$ ($\pi$) to obtain $\ket{\Phi^{(0)}}= \ket{\Phi^{+}}$ ($\ket{\Phi^{(\pi)}}= \ket{\Phi^{-}}$). Measured JSIs after the identity and Hadamard operations appear in Fig.~\ref{JSI}(b); in contrast to the $\ket{\Psi^\pm}$ case, positive frequency correlations are now clearly evident in the $\mathbb{I}_I\otimes\mathbb{I}_S$ measurement, with the Hadamard operation producing correlation patterns that depend on the state phase ($0$ or $\pi$).

Figure~\ref{JSI} reveals unique correlation signatures for each state in the results from both identity and Hadamard.
In fact, these signatures are sufficient to perform high-fidelity state quantum reconstruction
via Bayesian inference~\cite{lukens2020practical,lu2021full}, which has been shown to enable low-uncertainty estimates of highly correlated states measured in two pairs of MUBs~\cite{lu2018quantum, lu2022high}.
Our specific procedure~\cite{lu2021full} starts with a uniform (Bures) prior and uses a likelihood from the JSI measurements in the $Z\otimes Z$ and $X\otimes X$ bases. The estimated mean density matrices, shown in Fig.~\ref{Rho} have fidelities $\geq97\%$ with respect to the ideal Bell states. Interestingly, the fidelities for $\ket{\Psi^\pm}$ are slightly higher than those for $\ket{\Phi^\pm}$, which can be attributed to a combination of SPDC from residual $\omega_{P,0}$ pump and higher accidental coincidences in the latter case. In the dual-line pump scenario, the pump frequency at $\omega_{P,1}$ ($\omega_{P,-1}$) can also populate photons in frequency bins $I_0$ and $S_0$ ($I_1$ and $S_1)$ via downconversion in which the matched signal or idler falls outside of the computational space.
Such processes do not contribute to the ideal $\ket{\Phi^{\pm}}$ state in the coincidence basis. However, in practice the presence of multipair emission means that these undesired detection events can lead to accidental coincidences; specifically, for the same rate of desired coincidences for $\ket{\Psi^\pm}$ and $\ket{\Phi^\pm}$, the $\ket{\Phi^\pm}$ cases have double the rate of single-photon detection events, leading to a four-fold increase in uncorrelated coincidences. 

The impact of both imperfect carrier extinction and background processes are validated experimentally. First, for the $\mathbb{I}_I\otimes\mathbb{I}_S$ cases in Fig.~\ref{JSI}, 
the ratio of desired to undesired JSI points is around 50 for the $\ket{\Phi^\pm}$ states, in agreement with that predicted by the 17.5~dB carrier suppression in Fig.~\ref{setup}(a). (For the $\ket{\Psi^\pm}$ states where carrier suppression is not required, the mismatched JSI points fall to the observed accidentals level.) 
Similarly, computing the coincidences-to-accidentals ratios (CARs) found by comparing the coincidences in Fig.~\ref{JSI} against their values time-shifted in the raw histograms, we obtain CARs of $\sim$400   
for $\ket{\Psi^\pm}$ and $\sim$100  for $\ket{\Phi^\pm}$, again matching the four-fold theoretical prediction for accidentals between the two cases. We note that the first nonideality represents a technical limitation that could be eliminated with stronger suppression of the carrier frequency (through a different EOIM or additional pump filtering), whereas the second effect would require further engineering of the phase-matching function beyond the proposed configuration here. Nevertheless, the fidelities $\mathcal{F}\geq0.97$ which we have already observed---in the presence of these effects---highlight the immediate value of our approach for Bell state generation even without any further improvements.


\textit{Quantum delay sensing.---}The joint temporal correlation of time-energy entangled biphotons can be utilized for delay metrology with potential quantum advantages~\cite{giovannetti2004quantum,giovannetti2002positioning}. Negatively correlated entangled states (such as $\ket{\Psi^\pm}$) can probe changes to the difference in the delays traversed by the photons (differential-mode delay) via nonlocal measurements only~\cite{nonlocalquan2020high, seshadri2022nonlocal}. 
Entangled photons with positive frequency correlations (such as $\ket{\Phi^\pm}$) can offer enhancement in delay sensitivity beyond the shot noise limit~\cite{giovannetti2002positioning, kuzucu2005two, kuzucu2008joint}, by responding to changes in the sum of the signal and idler delays (common-mode delay). Combined, these complementary capabilities suggest that frequency-bin Bell states
could be employed for distributed sensing applications~\cite{zhao2021field,giovannetti2011advances} and monitoring delays and latencies in quantum networks. 
We highlight this potential using the demonstrated frequency-bin Bell states by examining their sensitivity to common-mode and differential-mode phase---fully equivalent to temporal delay through the general Fourier relationship between linear spectral phase and group delay~\cite{weiner2011ultrafast, pe2005temporal,seshadri2022nonlocal}. For the special case of two-dimensional systems, \emph{any} relative phase shift is trivially equal to a linear phase; thus any phase operation on a frequency-bin qubit can be mapped to a delay~\cite{lu2020fully}.
\begin{figure}[t]
  \centering
  \includegraphics[width=\linewidth]{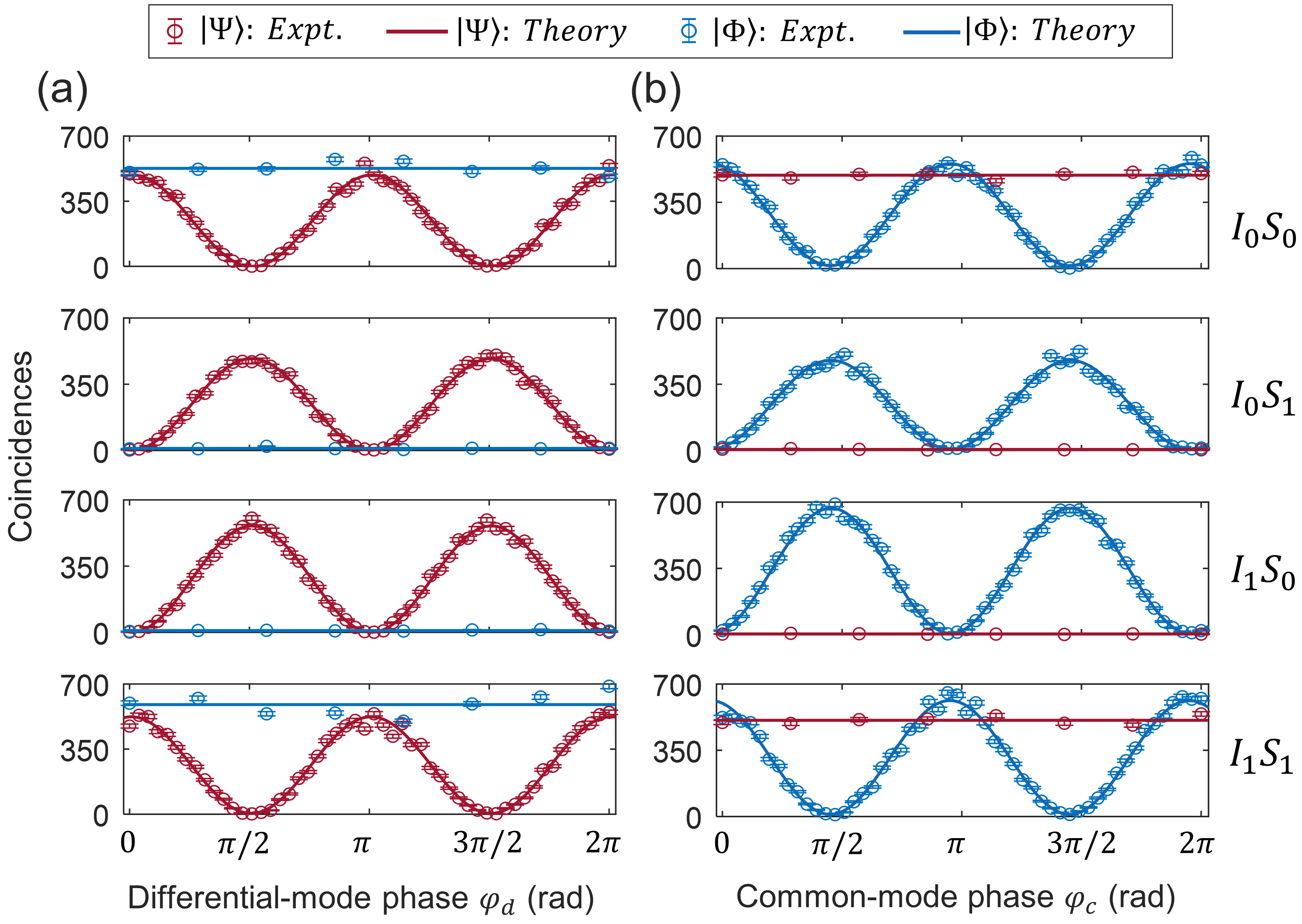}
\caption{Interferograms as (a)~differential-mode and (b)~common-mode spectral phases are scanned and the coincidences between different pairs of signal and idler bins are measured.}

\label{Interferograms}
\end{figure}

Specifically, a common-mode phase $\varphi_c$ applied on the bins $S_1$ and $I_1$  results in a global phase on the $\ket{\Psi^{\pm}}$ state which remains unaltered. However, the $\ket{\Phi^{+}}$ state transforms into
\begin{equation}\label{phi}
\ket{\Phi^{(2\varphi_c)}} = \ket{I_0S_0} + e^{2i\varphi_c}\ket{I_1S_1}.
\end{equation}
Such a transformation is equivalent to a common-mode delay of the form $\tau_c = (\tau_S + \tau_I)/2 = \varphi_c {\Delta\omega}^{-1}$, where $\tau_S$ ($\tau_I$) is the total delay experienced by the signal (idler) photon. That is, biphotons traveling through the same path accumulate phase corresponding to twice the delay traversed, the origin of quantum enhancement~\cite{giovannetti2002positioning}. 
After parallel Hadamard operations, the probability of coincidence detection between bins $I_0$ and $S_0$ becomes~\cite{Supplement}
\begin{equation}\label{ProbPhi}
\mathcal{P}^{X \otimes X}_{00}(\Phi^{(2\varphi_c)}) \propto \cos^2(2\varphi_c).
\end{equation}

On application of differential-mode phase $\varphi_d$ on the bins $S_0$ and $I_1$, the $\ket{\Phi^{\pm}}$ states are unaltered while $\ket{\Psi^{+}}$ transforms to
\begin{equation}
\ket{\Psi^{(2\varphi_d)}} = \ket{I_0S_1} + e^{2i\varphi_d}\ket{I_1S_0},
\end{equation}\label{psi}%
and the resultant coincidence probability between bins $I_0$ and $S_0$ after parallel Hadamard operations is 
\begin{equation}\label{ProbPsi}
\mathcal{P}^{X \otimes X}_{00}(\Psi^{(2\varphi_d)}) \propto \cos^2(2\varphi_d).
\end{equation}
This transformation is equivalent to a differential mode delay of the form $\tau_d = (\tau_S-\tau_I)/2 = {\varphi_d \Delta\omega}^{-1}$. 
The coincidence probabilities for other signal-idler bin pairs after the Hadamard operation are shown in Ref.~\cite{Supplement}. 
Using Shaper 1 to successively apply the common-mode phase ($\varphi_c$ on the bins $S_1$ and $I_1$) and differential-mode phase ($\varphi_d$ on the bins $S_0$ and $I_1$) prior to Hadamard operations, we measure coincidences between different frequency-bin pairs (plotted in Fig.~\ref{Interferograms}). The experiment clearly corroborates the theoretical prediction, highlighting the capability for differential- and common-mode delay metrology based on the frequency correlations of the probe state.

\textit{Discussion.---}We have demonstrated generation and tomography of all four two-qubit Bell states, to our knowledge the first time in frequency-bin encoding. Readily reconfigurable between single and dual line pump conditions and relying on passive spectral filtering, our setup can synthesize any Bell state within a fixed set of four frequency bins. The capability for on demand switching between the Bell pairs can find use in quantum cryptography applications~\cite{shi2013multi,chun2005secure}. Further, we demonstrate that the strong positive and negative frequency correlations in the generated Bell states can be used for sensing common-mode and differential-mode delays.


Moving forward, it will be interesting to explore the extent to which this technique can be generalized to higher-dimensional entanglement, the impact of coincidence basis postselection on protocols using these resources states, and methods to discriminate between them using frequency-bin Bell state analyzers~\cite{lingaraju2022bell}. This work also offers scope for analyzing related multi-line pump architectures for preparing two-qubit states. While on-chip generation of generic negatively correlated two-qudit states has been investigated~\cite{liscidini2019scalable}, our approach can offer further opportunities for on-chip implementation of $\ket{\Phi}$-like states utilizing multi-line pumped spontaneous four-wave mixing in single or series of microring resonantors (MRRs). In fact, through an appropriate cascade of MRR sources, it might be possible to suppress undesired biphoton generation processes such that only the signal and idler mode pairs of interest---i.e., those in the two-qubit computational basis---are efficiently produced. 

\begin{acknowledgments}
We thank AdvR for loaning the PPLN ridge waveguide. A portion of this work was performed at Oak Ridge National Laboratory, operated by UT-Battelle for the U.S. Department of Energy under contract no. DE-AC05-00OR22725. Funding was provided by the National Science Foundation (1839191-ECCS, 2034019-ECCS) and the U.S. Department of Energy, Office of Science, Advanced Scientific Computing Research, Early Career Research Program (Field Work Proposal ERKJ353).
\end{acknowledgments}

\section*{Supplement}
\label{theory}

\subsection*{Coincidence probability}

\begin{table*}
\centering
\begin{tabular}{llll}
\hline\rule{0pt}{1\normalbaselineskip}
Basis             &\hspace{7mm}Probability & \hspace{22mm}$\Tilde{\Psi}$ &\hspace{24mm} $\Tilde{\Phi}$  \\
\hline \rule{0pt}{1\normalbaselineskip}
\multirow{4}{*}{${Z\otimes Z}$} &   \hspace{12mm}$\mathcal{P}_{00}$         & \hspace{22.5mm}$0$ & \hspace{23mm}$\lvert \gamma_{00} \rvert^2$  \\
                  &\hspace{12mm}$\mathcal{P}_{01}$         &\hspace{21mm}$\lvert \gamma_{01}\rvert^2 $ &\hspace{25.5mm}$0$ \\
                  &\hspace{12mm}$\mathcal{P}_{10}$         & \hspace{21mm}$\lvert\gamma_{10}\rvert^2 $ &\hspace{25.5mm}$0$  \\
                  &  \hspace{12mm}$\mathcal{P}_{11}$        &\hspace{22.5mm}$0$ & \hspace{23mm}$\lvert\gamma_{11}\rvert^2$  \\[1.5mm]
                  
\multirow{2}{*}{${\hspace{1mm}X \otimes X}$} &  \hspace{8mm}$\mathcal{P}_{00} = \mathcal{P}_{11}$        & \hspace{8mm}$\eta^4\lvert \gamma_{01}e^{i\Delta\omega(\tau_S - \tau_I)} + \gamma_{10} \rvert^2$ & \hspace{8mm}$\eta^4\lvert \gamma_{00} + \gamma_{11}e^{i2\phi + i\Delta\omega(\tau_S + \tau_I)} \rvert^2$  \\
                  & \hspace{8mm}$\mathcal{P}_{01} = \mathcal{P}_{10}$           &\hspace{8mm}$\eta^4\lvert \gamma_{01}e^{i\Delta\omega(\tau_S - \tau_I)} - \gamma_{10} \rvert^2$  & \hspace{8mm}$\eta^4\lvert \gamma_{00} - \gamma_{11}e^{i2\phi + i\Delta\omega(\tau_S + \tau_I)} \rvert^2$\\
\hline
\end{tabular}
\caption{ The coincidence probabilities from ${Z \otimes Z}$ and ${X \otimes X}$ basis  measurements. }\label{ZXBasis}
\end{table*}

The negatively and positively correlated entangled states can be written as
\begin{equation}\label{Psi1}
\ket{\Tilde{\Psi}} = \Big(\gamma_{01} \hat{a}_{I,0}^\dagger \hat{a}_{S,1}^\dagger + \gamma_{10} \hat{a}_{I,1}^\dagger \hat{a}_{S,0}^\dagger\Big) \ket{\mathrm{vac}},
\end{equation} 
\begin{equation}\label{Phi1}
\ket{\Tilde{\Phi}} = \Big(\gamma_{00} \hat{a}_{I,0}^\dagger \hat{a}_{S,0}^\dagger + \gamma_{11} \hat{a}_{I,1}^\dagger \hat{a}_{S,1}^\dagger\Big) \ket{\mathrm{vac}},
\end{equation}
where $\ket{\mathrm{vac}}$ is the vacuum state, $\gamma_{kl}$ is the complex probability amplitude of the frequency bin pair associated with the creation operators $\hat{a}_{I,k}^\dagger$ and  $\hat{a}_{S,l}^\dagger$ corresponding to the $k^{th}$ idler bin ($I_k$) and $l^{th}$ signal bin ($S_l$) centered at $\omega_{I,k} = \frac{1}{2}\omega_{P,0} - \Omega_0 + (k-1)\Delta\omega $ and $\omega_{S,l}= \frac{1}{2}\omega_{P,0} + \Omega_0 + l\Delta\omega$ respectively, with $\Omega_0$ being the frequency offset of the bins $S_0$ and $I_1$ from the center frequency $\frac{1}{2}\omega_{p,0}$, and $\Delta\omega$ being the frequency-bin separation.

If the signal and idler traverse through delays given by $\tau_{_S}$ and $\tau_{_I}$ respectively (prior to the phase modulation illustrated in Fig.~1 of the main text), their annihilation operators transform into $\hat{b}_{I,k}$ and $\hat{b}_{S,l}$ as follows:
\begin{equation}\label{operatorI}
  \hat{b}_{I,k} = \hat{a}_{I,k} \exp{\big(i\tau_{_I}\omega_{{I,k}}\big)},
\end{equation}
\begin{equation}\label{operatorS}
  \hat{b}_{S,l} = \hat{a}_{S,l} \exp{\big(i\tau_{_S}\omega_{{S,l}}\big)}.
\end{equation}

Coincidence measurements between $k^{th}$ idler bin and $l^{th}$ signal bin in the $Z \otimes Z$ basis (i.e., when no phase modulation is applied) results in coincidence probabilities given by the following expressions and tabulated in Table~\ref{ZXBasis}:  
\begin{equation}\label{ProbZ}
\begin{aligned}
&\mathcal{P}^{Z \otimes Z}_{kl}(\Tilde{\Psi}) &= \Big|\bra{\mathrm{vac}} \hat{b}_{I,k} \hat{b}_{S,l} \ket{\Tilde{\Psi}}  \Big|^2\\
&\mathcal{P}^{Z \otimes Z}_{kl}(\Tilde{\Phi}) &= \Big|\bra{\mathrm{vac}} \hat{b}_{I,k} \hat{b}_{S,l} \ket{\Tilde{\Phi}}  \Big|^2\\
\end{aligned}
\end{equation}














On applying phase modulation of the form $\hat{m}(t) = e^{im\sin(\Delta\omega t + \phi)}$, the annihilation operators transform into
\begin{equation}
\begin{aligned}
\label{PhModI}
\hat{c}_{I,k} = \sum_{p=-1}^1 J_{p}(m) e^{-ip\phi} \hat{b}_{I,k-p}, 
\end{aligned}
\end{equation}
\begin{equation}
\begin{aligned}
\label{PhModS}
\hat{c}_{S,l} = \sum_{p=-1}^1 J_{p}(m) e^{-ip\phi} \hat{b}_{S,l-p}, 
\end{aligned}
\end{equation} where  $J_p(m)$ is the Bessel function of the first kind, $m$ is the modulation depth in radians, and $\phi$ is the phase of the RF sinusoidal waveform modulating the signal and idler photons. We limit the summation indices to $p\in\{-1,0,1\}$ to reflect the fact that only inputs in the qubit computational space are considered. On choosing the modulation index ($m = 1.435$ rad) such that $J_{1}(m) = J_0(m) = -J_{-1}(m) = 0.548 \equiv \eta$, the annihilation operators from Eqs.~(\ref{PhModI},\ref{PhModS}) can now be written as follows:

\begin{equation}
\begin{aligned}
\label{PhModI2}
\hat{c}_{I,k}(\phi) =  \eta \Big( \hat{b}_{I,k} + e^{-i\phi}\hat{b}_{I,k-1} - e^{i\phi}\hat{b}_{I,k+1}\Big), 
\end{aligned}
\end{equation}

\begin{equation}
\begin{aligned}
\label{PhModS2}
\hat{c}_{S,l}(\phi) = \eta \Big( \hat{b}_{S,l} + e^{-i\phi}\hat{b}_{S,l-1} - e^{i\phi}\hat{b}_{S,l+1}\Big). 
\end{aligned}
\end{equation}

Specifically, on setting the phase $\phi = 0$, Eqs.~(\ref{PhModI2},\ref{PhModS2}) essentially take the form of Hadamard operations 
as follows:


\begin{equation}
\begin{aligned}
\label{Hadamard_binNotation}
&\hat{c}_{I,0}(\phi=0) = \eta\left(\hat{b}_{I,0} -\hat{b}_{I,1}\right)\\
&\hat{c}_{S,0}(\phi=0) = \eta\left(\hat{b}_{S,0} - \hat{b}_{S,1}\right)\\
&\hat{c}_{I,1}(\phi=0) = \eta\left(\hat{b}_{I,1} + \hat{b}_{I,0}\right)\\
&\hat{c}_{S,1}(\phi=0) = \eta\left(\hat{b}_{S,1} + \hat{b}_{S,0}\right)\\
\end{aligned}
\end{equation}


On application of the transformation given by Eqs.~(\ref{PhModI2},\ref{PhModS2}) on the entangled states from Eqs.~(\ref{Psi1},\ref{Phi1}), the coincidence measurements between $k^{th}$ idler bin and $l^{th}$ signal bin in the $X \otimes X$ basis results in coincidence probabilities given by the following expressions and tabulated in Table~\ref{ZXBasis}:
\begin{equation}\label{ProbX}
\begin{aligned}
&\mathcal{P}^{X \otimes X}_{kl}(\Tilde{\Psi}) &= \Big|\bra{\mathrm{vac}} \hat{c}_{I,k} \hat{c}_{S,l} \ket{\Tilde{\Psi}}  \Big|^2\\
&\mathcal{P}^{X \otimes X}_{kl}(\Tilde{\Phi}) &= \Big|\bra{\mathrm{vac}} \hat{c}_{I,k} \hat{c}_{S,l} \ket{\Tilde{\Phi}}  \Big|^2\\
\end{aligned}
\end{equation}

\begin{widetext} 
\noindent For instance, $\mathcal{P}^{X \otimes X}_{00}(\Tilde{\Psi})$ can be computed as follows,

\begin{equation}\label{AlgProbX_00_Psi}
\begin{aligned}
&\mathcal{P}^{X \otimes X}_{00}(\Tilde{\Psi}) = \Big|\bra{\mathrm{vac}} \hat{c}_{I,0} \hat{c}_{S,0} \ket{\Tilde{\Psi}}  \Big|^2  \\
& = \eta^4\left| \left\langle{\mathrm{vac}} \left| \left( \hat{b}_{I,0} + e^{-i\phi}\hat{b}_{I,-1} -e^{i\phi}\hat{b}_{I,1}  \right)\left( \hat{b}_{S,0} + e^{-i\phi}\hat{b}_{S,-1} -e^{i\phi}\hat{b}_{S,1}  \right) \left( \gamma_{01}\hat{a}_{I,0}^\dagger\hat{a}_{S,1}^\dagger + \gamma_{10}\hat{a}_{I,1}^\dagger\hat{a}_{S,0}^\dagger \right) \right|{\mathrm{vac}} \right\rangle \right|^2 \\
& = \eta^4\left| \left\langle\mathrm{vac} \left|  \Big(e^{i\tau_{_I}\omega_{{I,0}}}\hat{a}_{I,0}\Big)\Big(-e^{i\phi}e^{i\tau_{_S}\omega_{{S,1}}}\hat{a}_{S,1}\Big)\gamma_{01}\hat{a}_{I,0}^\dagger\hat{a}_{S,1}^\dagger +\Big(-e^{i\phi}e^{i\tau_{_I}\omega_{{I,1}}}\hat{a}_{I,1}\Big) \Big(e^{i\tau_{_S}\omega_{{S,0}}}\hat{a}_{S,0} \Big)\gamma_{10}\hat{a}_{I,1}^\dagger\hat{a}_{S,0}^\dagger \right| \mathrm{ vac}\right\rangle  \right|^2\\
& = \eta^4\left| \gamma_{01}e^{i\Delta\omega(\tau_S - \tau_I)} + \gamma_{10} \right|^2 \\
\end{aligned}
\end{equation}{Similarly,}

\begin{equation}\label{AlgProbX_00_Phi}
\begin{aligned}
&\mathcal{P}^{X \otimes X}_{00}(\Tilde{\Phi}) = \Big|\bra{\mathrm{vac}} \hat{c}_{I,0} \hat{c}_{S,0} \ket{\Tilde{\Phi}}  \Big|^2  \\
& =  \eta^4 \left| \left\langle {\mathrm{vac}} \left| \left( \hat{b}_{I,0} + e^{-i\phi}\hat{b}_{I,-1} -e^{i\phi}\hat{b}_{I,1}  \right)\left( \hat{b}_{S,0} + e^{-i\phi}\hat{b}_{S,-1} -e^{i\phi}\hat{b}_{S,1}  \right) \left( \gamma_{00}\hat{a}_{I,0}^\dagger\hat{a}_{S,0}^\dagger + \gamma_{11}\hat{a}_{I,1}^\dagger\hat{a}_{S,1}^\dagger \right) \right| {\mathrm{ vac}} \right\rangle \right|^2 \\
& = \eta^4\left| \left\langle\mathrm{vac} \left| \Big(e^{i\tau_{_I}\omega_{{I,0}}}\hat{a}_{I,0}\Big)\Big(e^{i\tau_{_S}\omega_{{S,0}}}\hat{a}_{S,0}\Big)\gamma_{00}\hat{a}_{I,0}^\dagger\hat{a}_{S,0}^\dagger +\Big(-e^{i\phi}e^{i\tau_{_I}\omega_{{I,1}}}\hat{a}_{I,1}\Big) \Big(-e^{i\phi}e^{i\tau_{_S}\omega_{{S,1}}}\hat{a}_{S,1} \Big)\gamma_{11}\hat{a}_{I,1}^\dagger\hat{a}_{S,1}^\dagger \right| \mathrm{ vac}\right\rangle  \right|^2\\
& = \eta^4\left| \gamma_{00} + \gamma_{11}e^{i2\phi + i\Delta\omega(\tau_S + \tau_I)} \right|^2 \\
\end{aligned}
\end{equation}
\end{widetext}


We note that interference from the Hadamard operation reveals the phases in the frequency bin pairs constituting the Bell states. The negatively correlated $\ket{\Tilde{\Psi}}$ state is affected only by the difference in the delays traversed by the two photons while the positively correlated $\ket{\Tilde{\Phi}}$ state is affected only by the (common-mode) sum of the delays traversed by the two photons and equivalently the phase of the RF waveforms modulating the biphotons.
Specific to our experiment where the physical paths are such that $\tau_{_S} = \tau_{_I}$, we impart differential-mode and common-mode delays on a given state by applying phases using a pulse shaper, making use of the equivalence between group delay and linear spectral phase~\cite{pe2005temporal,seshadri2022nonlocal}.

\bibliography{refs}

\begin{thebibliography}{52}%
\makeatletter
\providecommand \@ifxundefined [1]{%
 \@ifx{#1\undefined}
}%
\providecommand \@ifnum [1]{%
 \ifnum #1\expandafter \@firstoftwo
 \else \expandafter \@secondoftwo
 \fi
}%
\providecommand \@ifx [1]{%
 \ifx #1\expandafter \@firstoftwo
 \else \expandafter \@secondoftwo
 \fi
}%
\providecommand \natexlab [1]{#1}%
\providecommand \enquote  [1]{``#1''}%
\providecommand \bibnamefont  [1]{#1}%
\providecommand \bibfnamefont [1]{#1}%
\providecommand \citenamefont [1]{#1}%
\providecommand \href@noop [0]{\@secondoftwo}%
\providecommand \href [0]{\begingroup \@sanitize@url \@href}%
\providecommand \@href[1]{\@@startlink{#1}\@@href}%
\providecommand \@@href[1]{\endgroup#1\@@endlink}%
\providecommand \@sanitize@url [0]{\catcode `\\12\catcode `\$12\catcode
  `\&12\catcode `\#12\catcode `\^12\catcode `\_12\catcode `\%12\relax}%
\providecommand \@@startlink[1]{}%
\providecommand \@@endlink[0]{}%
\providecommand \url  [0]{\begingroup\@sanitize@url \@url }%
\providecommand \@url [1]{\endgroup\@href {#1}{\urlprefix }}%
\providecommand \urlprefix  [0]{URL }%
\providecommand \Eprint [0]{\href }%
\providecommand \doibase [0]{http://dx.doi.org/}%
\providecommand \selectlanguage [0]{\@gobble}%
\providecommand \bibinfo  [0]{\@secondoftwo}%
\providecommand \bibfield  [0]{\@secondoftwo}%
\providecommand \translation [1]{[#1]}%
\providecommand \BibitemOpen [0]{}%
\providecommand \bibitemStop [0]{}%
\providecommand \bibitemNoStop [0]{.\EOS\space}%
\providecommand \EOS [0]{\spacefactor3000\relax}%
\providecommand \BibitemShut  [1]{\csname bibitem#1\endcsname}%
\let\auto@bib@innerbib\@empty
\bibitem [{\citenamefont {Mattle}\ \emph {et~al.}(1996)\citenamefont {Mattle},
  \citenamefont {Weinfurter}, \citenamefont {Kwiat},\ and\ \citenamefont
  {Zeilinger}}]{mattle1996dense}%
  \BibitemOpen
  \bibfield  {author} {\bibinfo {author} {\bibfnamefont {K.}~\bibnamefont
  {Mattle}}, \bibinfo {author} {\bibfnamefont {H.}~\bibnamefont {Weinfurter}},
  \bibinfo {author} {\bibfnamefont {P.~G.}\ \bibnamefont {Kwiat}}, \ and\
  \bibinfo {author} {\bibfnamefont {A.}~\bibnamefont {Zeilinger}},\ }\href@noop
  {} {\bibfield  {journal} {\bibinfo  {journal} {Phys. Rev. Lett.}\ }\textbf
  {\bibinfo {volume} {76}},\ \bibinfo {pages} {4656} (\bibinfo {year}
  {1996})}\BibitemShut {NoStop}%
\bibitem [{\citenamefont {Bennett}\ \emph {et~al.}(1993)\citenamefont
  {Bennett}, \citenamefont {Brassard}, \citenamefont {Cr{\'e}peau},
  \citenamefont {Jozsa}, \citenamefont {Peres},\ and\ \citenamefont
  {Wootters}}]{bennett1993teleporting}%
  \BibitemOpen
  \bibfield  {author} {\bibinfo {author} {\bibfnamefont {C.~H.}\ \bibnamefont
  {Bennett}}, \bibinfo {author} {\bibfnamefont {G.}~\bibnamefont {Brassard}},
  \bibinfo {author} {\bibfnamefont {C.}~\bibnamefont {Cr{\'e}peau}}, \bibinfo
  {author} {\bibfnamefont {R.}~\bibnamefont {Jozsa}}, \bibinfo {author}
  {\bibfnamefont {A.}~\bibnamefont {Peres}}, \ and\ \bibinfo {author}
  {\bibfnamefont {W.~K.}\ \bibnamefont {Wootters}},\ }\href@noop {} {\bibfield
  {journal} {\bibinfo  {journal} {Phys. Rev. Lett.}\ }\textbf {\bibinfo
  {volume} {70}},\ \bibinfo {pages} {1895} (\bibinfo {year}
  {1993})}\BibitemShut {NoStop}%
\bibitem [{\citenamefont {Shukla}\ \emph {et~al.}(2014)\citenamefont {Shukla},
  \citenamefont {Alam},\ and\ \citenamefont {Pathak}}]{shukla2014protocols}%
  \BibitemOpen
  \bibfield  {author} {\bibinfo {author} {\bibfnamefont {C.}~\bibnamefont
  {Shukla}}, \bibinfo {author} {\bibfnamefont {N.}~\bibnamefont {Alam}}, \ and\
  \bibinfo {author} {\bibfnamefont {A.}~\bibnamefont {Pathak}},\ }\href@noop {}
  {\bibfield  {journal} {\bibinfo  {journal} {Quantum Inf. Process.}\ }\textbf
  {\bibinfo {volume} {13}},\ \bibinfo {pages} {2391} (\bibinfo {year}
  {2014})}\BibitemShut {NoStop}%
\bibitem [{\citenamefont {Shi}\ and\ \citenamefont
  {Zhong}(2013)}]{shi2013multi}%
  \BibitemOpen
  \bibfield  {author} {\bibinfo {author} {\bibfnamefont {R.-H.}\ \bibnamefont
  {Shi}}\ and\ \bibinfo {author} {\bibfnamefont {H.}~\bibnamefont {Zhong}},\
  }\href@noop {} {\bibfield  {journal} {\bibinfo  {journal} {Quantum Inf.
  Process.}\ }\textbf {\bibinfo {volume} {12}},\ \bibinfo {pages} {921}
  (\bibinfo {year} {2013})}\BibitemShut {NoStop}%
\bibitem [{\citenamefont {Tittel}\ \emph {et~al.}(2000)\citenamefont {Tittel},
  \citenamefont {Brendel}, \citenamefont {Zbinden},\ and\ \citenamefont
  {Gisin}}]{tittel2000quantum}%
  \BibitemOpen
  \bibfield  {author} {\bibinfo {author} {\bibfnamefont {W.}~\bibnamefont
  {Tittel}}, \bibinfo {author} {\bibfnamefont {J.}~\bibnamefont {Brendel}},
  \bibinfo {author} {\bibfnamefont {H.}~\bibnamefont {Zbinden}}, \ and\
  \bibinfo {author} {\bibfnamefont {N.}~\bibnamefont {Gisin}},\ }\href@noop {}
  {\bibfield  {journal} {\bibinfo  {journal} {Phys. Rev. Lett.}\ }\textbf
  {\bibinfo {volume} {84}},\ \bibinfo {pages} {4737} (\bibinfo {year}
  {2000})}\BibitemShut {NoStop}%
\bibitem [{\citenamefont {Pan}\ \emph {et~al.}(1998)\citenamefont {Pan},
  \citenamefont {Bouwmeester}, \citenamefont {Weinfurter},\ and\ \citenamefont
  {Zeilinger}}]{pan1998experimental}%
  \BibitemOpen
  \bibfield  {author} {\bibinfo {author} {\bibfnamefont {J.-W.}\ \bibnamefont
  {Pan}}, \bibinfo {author} {\bibfnamefont {D.}~\bibnamefont {Bouwmeester}},
  \bibinfo {author} {\bibfnamefont {H.}~\bibnamefont {Weinfurter}}, \ and\
  \bibinfo {author} {\bibfnamefont {A.}~\bibnamefont {Zeilinger}},\ }\href@noop
  {} {\bibfield  {journal} {\bibinfo  {journal} {Phys. Rev. Lett.}\ }\textbf
  {\bibinfo {volume} {80}},\ \bibinfo {pages} {3891} (\bibinfo {year}
  {1998})}\BibitemShut {NoStop}%
\bibitem [{\citenamefont {Goebel}\ \emph {et~al.}(2008)\citenamefont {Goebel},
  \citenamefont {Wagenknecht}, \citenamefont {Zhang}, \citenamefont {Chen},
  \citenamefont {Chen}, \citenamefont {Schmiedmayer},\ and\ \citenamefont
  {Pan}}]{goebel2008multistage}%
  \BibitemOpen
  \bibfield  {author} {\bibinfo {author} {\bibfnamefont {A.~M.}\ \bibnamefont
  {Goebel}}, \bibinfo {author} {\bibfnamefont {C.}~\bibnamefont {Wagenknecht}},
  \bibinfo {author} {\bibfnamefont {Q.}~\bibnamefont {Zhang}}, \bibinfo
  {author} {\bibfnamefont {Y.-A.}\ \bibnamefont {Chen}}, \bibinfo {author}
  {\bibfnamefont {K.}~\bibnamefont {Chen}}, \bibinfo {author} {\bibfnamefont
  {J.}~\bibnamefont {Schmiedmayer}}, \ and\ \bibinfo {author} {\bibfnamefont
  {J.-W.}\ \bibnamefont {Pan}},\ }\href@noop {} {\bibfield  {journal} {\bibinfo
   {journal} {Phys. Rev. Lett.}\ }\textbf {\bibinfo {volume} {101}},\ \bibinfo
  {pages} {080403} (\bibinfo {year} {2008})}\BibitemShut {NoStop}%
\bibitem [{\citenamefont {Kwiat}\ \emph {et~al.}(1995)\citenamefont {Kwiat},
  \citenamefont {Mattle}, \citenamefont {Weinfurter}, \citenamefont
  {Zeilinger}, \citenamefont {Sergienko},\ and\ \citenamefont
  {Shih}}]{kwiat1995new}%
  \BibitemOpen
  \bibfield  {author} {\bibinfo {author} {\bibfnamefont {P.~G.}\ \bibnamefont
  {Kwiat}}, \bibinfo {author} {\bibfnamefont {K.}~\bibnamefont {Mattle}},
  \bibinfo {author} {\bibfnamefont {H.}~\bibnamefont {Weinfurter}}, \bibinfo
  {author} {\bibfnamefont {A.}~\bibnamefont {Zeilinger}}, \bibinfo {author}
  {\bibfnamefont {A.~V.}\ \bibnamefont {Sergienko}}, \ and\ \bibinfo {author}
  {\bibfnamefont {Y.}~\bibnamefont {Shih}},\ }\href@noop {} {\bibfield
  {journal} {\bibinfo  {journal} {Phys. Rev. Lett.}\ }\textbf {\bibinfo
  {volume} {75}},\ \bibinfo {pages} {4337} (\bibinfo {year}
  {1995})}\BibitemShut {NoStop}%
\bibitem [{\citenamefont {Agnew}\ \emph {et~al.}(2013)\citenamefont {Agnew},
  \citenamefont {Salvail}, \citenamefont {Leach},\ and\ \citenamefont
  {Boyd}}]{agnew2013generation}%
  \BibitemOpen
  \bibfield  {author} {\bibinfo {author} {\bibfnamefont {M.}~\bibnamefont
  {Agnew}}, \bibinfo {author} {\bibfnamefont {J.~Z.}\ \bibnamefont {Salvail}},
  \bibinfo {author} {\bibfnamefont {J.}~\bibnamefont {Leach}}, \ and\ \bibinfo
  {author} {\bibfnamefont {R.~W.}\ \bibnamefont {Boyd}},\ }\href@noop {}
  {\bibfield  {journal} {\bibinfo  {journal} {Phys. Rev. Lett.}\ }\textbf
  {\bibinfo {volume} {111}},\ \bibinfo {pages} {030402} (\bibinfo {year}
  {2013})}\BibitemShut {NoStop}%
\bibitem [{\citenamefont {Lo}\ \emph {et~al.}(2020)\citenamefont {Lo},
  \citenamefont {Ikuta}, \citenamefont {Matsuda}, \citenamefont {Honjo},
  \citenamefont {Munro},\ and\ \citenamefont {Takesue}}]{Lo2020}%
  \BibitemOpen
  \bibfield  {author} {\bibinfo {author} {\bibfnamefont {H.-P.}\ \bibnamefont
  {Lo}}, \bibinfo {author} {\bibfnamefont {T.}~\bibnamefont {Ikuta}}, \bibinfo
  {author} {\bibfnamefont {N.}~\bibnamefont {Matsuda}}, \bibinfo {author}
  {\bibfnamefont {T.}~\bibnamefont {Honjo}}, \bibinfo {author} {\bibfnamefont
  {W.~J.}\ \bibnamefont {Munro}}, \ and\ \bibinfo {author} {\bibfnamefont
  {H.}~\bibnamefont {Takesue}},\ }\href {\doibase
  10.1103/PhysRevApplied.13.034013} {\bibfield  {journal} {\bibinfo  {journal}
  {Phys. Rev. Applied}\ }\textbf {\bibinfo {volume} {13}},\ \bibinfo {pages}
  {034013} (\bibinfo {year} {2020})}\BibitemShut {NoStop}%
\bibitem [{\citenamefont {Brendel}\ \emph {et~al.}(1999)\citenamefont
  {Brendel}, \citenamefont {Gisin}, \citenamefont {Tittel},\ and\ \citenamefont
  {Zbinden}}]{brendel1999pulsed}%
  \BibitemOpen
  \bibfield  {author} {\bibinfo {author} {\bibfnamefont {J.}~\bibnamefont
  {Brendel}}, \bibinfo {author} {\bibfnamefont {N.}~\bibnamefont {Gisin}},
  \bibinfo {author} {\bibfnamefont {W.}~\bibnamefont {Tittel}}, \ and\ \bibinfo
  {author} {\bibfnamefont {H.}~\bibnamefont {Zbinden}},\ }\href@noop {}
  {\bibfield  {journal} {\bibinfo  {journal} {Phys. Rev. Lett.}\ }\textbf
  {\bibinfo {volume} {82}},\ \bibinfo {pages} {2594} (\bibinfo {year}
  {1999})}\BibitemShut {NoStop}%
\bibitem [{\citenamefont {Shadbolt}\ \emph {et~al.}(2012)\citenamefont
  {Shadbolt}, \citenamefont {Verde}, \citenamefont {Peruzzo}, \citenamefont
  {Politi}, \citenamefont {Laing}, \citenamefont {Lobino}, \citenamefont
  {Matthews}, \citenamefont {Thompson},\ and\ \citenamefont
  {O'Brien}}]{shadbolt2012generating}%
  \BibitemOpen
  \bibfield  {author} {\bibinfo {author} {\bibfnamefont {P.~J.}\ \bibnamefont
  {Shadbolt}}, \bibinfo {author} {\bibfnamefont {M.~R.}\ \bibnamefont {Verde}},
  \bibinfo {author} {\bibfnamefont {A.}~\bibnamefont {Peruzzo}}, \bibinfo
  {author} {\bibfnamefont {A.}~\bibnamefont {Politi}}, \bibinfo {author}
  {\bibfnamefont {A.}~\bibnamefont {Laing}}, \bibinfo {author} {\bibfnamefont
  {M.}~\bibnamefont {Lobino}}, \bibinfo {author} {\bibfnamefont {J.~C.}\
  \bibnamefont {Matthews}}, \bibinfo {author} {\bibfnamefont {M.~G.}\
  \bibnamefont {Thompson}}, \ and\ \bibinfo {author} {\bibfnamefont {J.~L.}\
  \bibnamefont {O'Brien}},\ }\href@noop {} {\bibfield  {journal} {\bibinfo
  {journal} {Nat. Photon.}\ }\textbf {\bibinfo {volume} {6}},\ \bibinfo {pages}
  {45} (\bibinfo {year} {2012})}\BibitemShut {NoStop}%
\bibitem [{\citenamefont {Silverstone}\ \emph {et~al.}(2014)\citenamefont
  {Silverstone}, \citenamefont {Bonneau}, \citenamefont {Ohira}, \citenamefont
  {Suzuki}, \citenamefont {Yoshida}, \citenamefont {Iizuka}, \citenamefont
  {Ezaki}, \citenamefont {Natarajan}, \citenamefont {Tanner}, \citenamefont
  {Hadfield}, \citenamefont {Zwiller}, \citenamefont {Marshall}, \citenamefont
  {Rarity}, \citenamefont {O'Brien},\ and\ \citenamefont
  {Thompson}}]{silverstone2014chip}%
  \BibitemOpen
  \bibfield  {author} {\bibinfo {author} {\bibfnamefont {J.~W.}\ \bibnamefont
  {Silverstone}}, \bibinfo {author} {\bibfnamefont {D.}~\bibnamefont
  {Bonneau}}, \bibinfo {author} {\bibfnamefont {K.}~\bibnamefont {Ohira}},
  \bibinfo {author} {\bibfnamefont {N.}~\bibnamefont {Suzuki}}, \bibinfo
  {author} {\bibfnamefont {H.}~\bibnamefont {Yoshida}}, \bibinfo {author}
  {\bibfnamefont {N.}~\bibnamefont {Iizuka}}, \bibinfo {author} {\bibfnamefont
  {M.}~\bibnamefont {Ezaki}}, \bibinfo {author} {\bibfnamefont {C.~M.}\
  \bibnamefont {Natarajan}}, \bibinfo {author} {\bibfnamefont {M.~G.}\
  \bibnamefont {Tanner}}, \bibinfo {author} {\bibfnamefont {R.~H.}\
  \bibnamefont {Hadfield}}, \bibinfo {author} {\bibfnamefont {V.}~\bibnamefont
  {Zwiller}}, \bibinfo {author} {\bibfnamefont {G.~D.}\ \bibnamefont
  {Marshall}}, \bibinfo {author} {\bibfnamefont {J.~G.}\ \bibnamefont
  {Rarity}}, \bibinfo {author} {\bibfnamefont {J.~L.}\ \bibnamefont {O'Brien}},
  \ and\ \bibinfo {author} {\bibfnamefont {M.~G.}\ \bibnamefont {Thompson}},\
  }\href {\doibase 10.1038/nphoton.2013.339} {\bibfield  {journal} {\bibinfo
  {journal} {Nat. Photon.}\ }\textbf {\bibinfo {volume} {8}},\ \bibinfo {pages}
  {104} (\bibinfo {year} {2014})}\BibitemShut {NoStop}%
\bibitem [{\citenamefont {Li}\ \emph {et~al.}(2020)\citenamefont {Li},
  \citenamefont {Zhang}, \citenamefont {Chen}, \citenamefont {Ren},
  \citenamefont {Gong},\ and\ \citenamefont {Li}}]{li2020femtosecond}%
  \BibitemOpen
  \bibfield  {author} {\bibinfo {author} {\bibfnamefont {M.}~\bibnamefont
  {Li}}, \bibinfo {author} {\bibfnamefont {Q.}~\bibnamefont {Zhang}}, \bibinfo
  {author} {\bibfnamefont {Y.}~\bibnamefont {Chen}}, \bibinfo {author}
  {\bibfnamefont {X.}~\bibnamefont {Ren}}, \bibinfo {author} {\bibfnamefont
  {Q.}~\bibnamefont {Gong}}, \ and\ \bibinfo {author} {\bibfnamefont
  {Y.}~\bibnamefont {Li}},\ }\href@noop {} {\bibfield  {journal} {\bibinfo
  {journal} {Micromachines}\ }\textbf {\bibinfo {volume} {11}},\ \bibinfo
  {pages} {1111} (\bibinfo {year} {2020})}\BibitemShut {NoStop}%
\bibitem [{\citenamefont {Kues}\ \emph {et~al.}(2019)\citenamefont {Kues},
  \citenamefont {Reimer}, \citenamefont {Lukens}, \citenamefont {Munro},
  \citenamefont {Weiner}, \citenamefont {Moss},\ and\ \citenamefont
  {Morandotti}}]{kues2019quantum}%
  \BibitemOpen
  \bibfield  {author} {\bibinfo {author} {\bibfnamefont {M.}~\bibnamefont
  {Kues}}, \bibinfo {author} {\bibfnamefont {C.}~\bibnamefont {Reimer}},
  \bibinfo {author} {\bibfnamefont {J.~M.}\ \bibnamefont {Lukens}}, \bibinfo
  {author} {\bibfnamefont {W.~J.}\ \bibnamefont {Munro}}, \bibinfo {author}
  {\bibfnamefont {A.~M.}\ \bibnamefont {Weiner}}, \bibinfo {author}
  {\bibfnamefont {D.~J.}\ \bibnamefont {Moss}}, \ and\ \bibinfo {author}
  {\bibfnamefont {R.}~\bibnamefont {Morandotti}},\ }\href@noop {} {\bibfield
  {journal} {\bibinfo  {journal} {Nat. Photon.}\ }\textbf {\bibinfo {volume}
  {13}},\ \bibinfo {pages} {170} (\bibinfo {year} {2019})}\BibitemShut
  {NoStop}%
\bibitem [{\citenamefont {Imany}\ \emph
  {et~al.}(2018{\natexlab{a}})\citenamefont {Imany}, \citenamefont
  {Jaramillo-Villegas}, \citenamefont {Odele}, \citenamefont {Han},
  \citenamefont {Leaird}, \citenamefont {Lukens}, \citenamefont {Lougovski},
  \citenamefont {Qi},\ and\ \citenamefont {Weiner}}]{imany201850}%
  \BibitemOpen
  \bibfield  {author} {\bibinfo {author} {\bibfnamefont {P.}~\bibnamefont
  {Imany}}, \bibinfo {author} {\bibfnamefont {J.~A.}\ \bibnamefont
  {Jaramillo-Villegas}}, \bibinfo {author} {\bibfnamefont {O.~D.}\ \bibnamefont
  {Odele}}, \bibinfo {author} {\bibfnamefont {K.}~\bibnamefont {Han}}, \bibinfo
  {author} {\bibfnamefont {D.~E.}\ \bibnamefont {Leaird}}, \bibinfo {author}
  {\bibfnamefont {J.~M.}\ \bibnamefont {Lukens}}, \bibinfo {author}
  {\bibfnamefont {P.}~\bibnamefont {Lougovski}}, \bibinfo {author}
  {\bibfnamefont {M.}~\bibnamefont {Qi}}, \ and\ \bibinfo {author}
  {\bibfnamefont {A.~M.}\ \bibnamefont {Weiner}},\ }\href@noop {} {\bibfield
  {journal} {\bibinfo  {journal} {Opt. Express}\ }\textbf {\bibinfo {volume}
  {26}},\ \bibinfo {pages} {1825} (\bibinfo {year}
  {2018}{\natexlab{a}})}\BibitemShut {NoStop}%
\bibitem [{\citenamefont {Lu}\ \emph {et~al.}(2018{\natexlab{a}})\citenamefont
  {Lu}, \citenamefont {Lukens}, \citenamefont {Peters}, \citenamefont {Odele},
  \citenamefont {Leaird}, \citenamefont {Weiner},\ and\ \citenamefont
  {Lougovski}}]{lu2018electro}%
  \BibitemOpen
  \bibfield  {author} {\bibinfo {author} {\bibfnamefont {H.-H.}\ \bibnamefont
  {Lu}}, \bibinfo {author} {\bibfnamefont {J.~M.}\ \bibnamefont {Lukens}},
  \bibinfo {author} {\bibfnamefont {N.~A.}\ \bibnamefont {Peters}}, \bibinfo
  {author} {\bibfnamefont {O.~D.}\ \bibnamefont {Odele}}, \bibinfo {author}
  {\bibfnamefont {D.~E.}\ \bibnamefont {Leaird}}, \bibinfo {author}
  {\bibfnamefont {A.~M.}\ \bibnamefont {Weiner}}, \ and\ \bibinfo {author}
  {\bibfnamefont {P.}~\bibnamefont {Lougovski}},\ }\href@noop {} {\bibfield
  {journal} {\bibinfo  {journal} {Phys. Rev. Lett.}\ }\textbf {\bibinfo
  {volume} {120}},\ \bibinfo {pages} {030502} (\bibinfo {year}
  {2018}{\natexlab{a}})}\BibitemShut {NoStop}%
\bibitem [{\citenamefont {Zhang}\ \emph {et~al.}(2021)\citenamefont {Zhang},
  \citenamefont {Cui}, \citenamefont {Yan}, \citenamefont {Guo}, \citenamefont
  {Wang},\ and\ \citenamefont {Fan}}]{zhang2021chip}%
  \BibitemOpen
  \bibfield  {author} {\bibinfo {author} {\bibfnamefont {L.}~\bibnamefont
  {Zhang}}, \bibinfo {author} {\bibfnamefont {C.}~\bibnamefont {Cui}}, \bibinfo
  {author} {\bibfnamefont {J.}~\bibnamefont {Yan}}, \bibinfo {author}
  {\bibfnamefont {Y.}~\bibnamefont {Guo}}, \bibinfo {author} {\bibfnamefont
  {J.}~\bibnamefont {Wang}}, \ and\ \bibinfo {author} {\bibfnamefont
  {L.}~\bibnamefont {Fan}},\ }\href@noop {} {\bibfield  {journal} {\bibinfo
  {journal} {arXiv:2111.12784}\ } (\bibinfo {year} {2021})}\BibitemShut
  {NoStop}%
\bibitem [{\citenamefont {Lu}\ \emph {et~al.}(2018{\natexlab{b}})\citenamefont
  {Lu}, \citenamefont {Lukens}, \citenamefont {Peters}, \citenamefont
  {Williams}, \citenamefont {Weiner},\ and\ \citenamefont
  {Lougovski}}]{lu2018quantum}%
  \BibitemOpen
  \bibfield  {author} {\bibinfo {author} {\bibfnamefont {H.-H.}\ \bibnamefont
  {Lu}}, \bibinfo {author} {\bibfnamefont {J.~M.}\ \bibnamefont {Lukens}},
  \bibinfo {author} {\bibfnamefont {N.~A.}\ \bibnamefont {Peters}}, \bibinfo
  {author} {\bibfnamefont {B.~P.}\ \bibnamefont {Williams}}, \bibinfo {author}
  {\bibfnamefont {A.~M.}\ \bibnamefont {Weiner}}, \ and\ \bibinfo {author}
  {\bibfnamefont {P.}~\bibnamefont {Lougovski}},\ }\href@noop {} {\bibfield
  {journal} {\bibinfo  {journal} {Optica}\ }\textbf {\bibinfo {volume} {5}},\
  \bibinfo {pages} {1455} (\bibinfo {year} {2018}{\natexlab{b}})}\BibitemShut
  {NoStop}%
\bibitem [{\citenamefont {Lingaraju}\ \emph {et~al.}(2022)\citenamefont
  {Lingaraju}, \citenamefont {Lu}, \citenamefont {Leaird}, \citenamefont
  {Estrella}, \citenamefont {Lukens},\ and\ \citenamefont
  {Weiner}}]{lingaraju2022bell}%
  \BibitemOpen
  \bibfield  {author} {\bibinfo {author} {\bibfnamefont {N.~B.}\ \bibnamefont
  {Lingaraju}}, \bibinfo {author} {\bibfnamefont {H.-H.}\ \bibnamefont {Lu}},
  \bibinfo {author} {\bibfnamefont {D.~E.}\ \bibnamefont {Leaird}}, \bibinfo
  {author} {\bibfnamefont {S.}~\bibnamefont {Estrella}}, \bibinfo {author}
  {\bibfnamefont {J.~M.}\ \bibnamefont {Lukens}}, \ and\ \bibinfo {author}
  {\bibfnamefont {A.~M.}\ \bibnamefont {Weiner}},\ }\href@noop {} {\bibfield
  {journal} {\bibinfo  {journal} {Optica}\ }\textbf {\bibinfo {volume} {9}},\
  \bibinfo {pages} {280} (\bibinfo {year} {2022})}\BibitemShut {NoStop}%
\bibitem [{\citenamefont {Lu}\ \emph {et~al.}(2020)\citenamefont {Lu},
  \citenamefont {Simmerman}, \citenamefont {Lougovski}, \citenamefont
  {Weiner},\ and\ \citenamefont {Lukens}}]{lu2020fully}%
  \BibitemOpen
  \bibfield  {author} {\bibinfo {author} {\bibfnamefont {H.-H.}\ \bibnamefont
  {Lu}}, \bibinfo {author} {\bibfnamefont {E.~M.}\ \bibnamefont {Simmerman}},
  \bibinfo {author} {\bibfnamefont {P.}~\bibnamefont {Lougovski}}, \bibinfo
  {author} {\bibfnamefont {A.~M.}\ \bibnamefont {Weiner}}, \ and\ \bibinfo
  {author} {\bibfnamefont {J.~M.}\ \bibnamefont {Lukens}},\ }\href@noop {}
  {\bibfield  {journal} {\bibinfo  {journal} {Phys. Rev. Lett.}\ }\textbf
  {\bibinfo {volume} {125}},\ \bibinfo {pages} {120503} (\bibinfo {year}
  {2020})}\BibitemShut {NoStop}%
\bibitem [{\citenamefont {Giovannetti}\ \emph {et~al.}(2002)\citenamefont
  {Giovannetti}, \citenamefont {Lloyd},\ and\ \citenamefont
  {Maccone}}]{giovannetti2002positioning}%
  \BibitemOpen
  \bibfield  {author} {\bibinfo {author} {\bibfnamefont {V.}~\bibnamefont
  {Giovannetti}}, \bibinfo {author} {\bibfnamefont {S.}~\bibnamefont {Lloyd}},
  \ and\ \bibinfo {author} {\bibfnamefont {L.}~\bibnamefont {Maccone}},\
  }\href@noop {} {\bibfield  {journal} {\bibinfo  {journal} {Phys. Rev. A}\
  }\textbf {\bibinfo {volume} {65}},\ \bibinfo {pages} {022309} (\bibinfo
  {year} {2002})}\BibitemShut {NoStop}%
\bibitem [{\citenamefont {Giovannetti}\ \emph {et~al.}(2001)\citenamefont
  {Giovannetti}, \citenamefont {Lloyd},\ and\ \citenamefont
  {Maccone}}]{giovannetti2001quantum}%
  \BibitemOpen
  \bibfield  {author} {\bibinfo {author} {\bibfnamefont {V.}~\bibnamefont
  {Giovannetti}}, \bibinfo {author} {\bibfnamefont {S.}~\bibnamefont {Lloyd}},
  \ and\ \bibinfo {author} {\bibfnamefont {L.}~\bibnamefont {Maccone}},\
  }\href@noop {} {\bibfield  {journal} {\bibinfo  {journal} {Nature}\ }\textbf
  {\bibinfo {volume} {412}},\ \bibinfo {pages} {417} (\bibinfo {year}
  {2001})}\BibitemShut {NoStop}%
\bibitem [{\citenamefont {Grice}\ and\ \citenamefont
  {Walmsley}(1997)}]{grice1997spectral}%
  \BibitemOpen
  \bibfield  {author} {\bibinfo {author} {\bibfnamefont {W.~P.}\ \bibnamefont
  {Grice}}\ and\ \bibinfo {author} {\bibfnamefont {I.~A.}\ \bibnamefont
  {Walmsley}},\ }\href@noop {} {\bibfield  {journal} {\bibinfo  {journal}
  {Phys. Rev. A}\ }\textbf {\bibinfo {volume} {56}},\ \bibinfo {pages} {1627}
  (\bibinfo {year} {1997})}\BibitemShut {NoStop}%
\bibitem [{\citenamefont {Hum}\ and\ \citenamefont
  {Fejer}(2007)}]{hum2007quasi}%
  \BibitemOpen
  \bibfield  {author} {\bibinfo {author} {\bibfnamefont {D.~S.}\ \bibnamefont
  {Hum}}\ and\ \bibinfo {author} {\bibfnamefont {M.~M.}\ \bibnamefont
  {Fejer}},\ }\href@noop {} {\bibfield  {journal} {\bibinfo  {journal} {Comptes
  Rendus Physique}\ }\textbf {\bibinfo {volume} {8}},\ \bibinfo {pages} {180}
  (\bibinfo {year} {2007})}\BibitemShut {NoStop}%
\bibitem [{\citenamefont {Harris}(2007)}]{Harris2007}%
  \BibitemOpen
  \bibfield  {author} {\bibinfo {author} {\bibfnamefont {S.~E.}\ \bibnamefont
  {Harris}},\ }\href {\doibase 10.1103/PhysRevLett.98.063602} {\bibfield
  {journal} {\bibinfo  {journal} {Phys. Rev. Lett.}\ }\textbf {\bibinfo
  {volume} {98}},\ \bibinfo {pages} {063602} (\bibinfo {year}
  {2007})}\BibitemShut {NoStop}%
\bibitem [{\citenamefont {Nasr}\ \emph {et~al.}(2008)\citenamefont {Nasr},
  \citenamefont {Carrasco}, \citenamefont {Saleh}, \citenamefont {Sergienko},
  \citenamefont {Teich}, \citenamefont {Torres}, \citenamefont {Torner},
  \citenamefont {Hum},\ and\ \citenamefont {Fejer}}]{Nasr2008}%
  \BibitemOpen
  \bibfield  {author} {\bibinfo {author} {\bibfnamefont {M.~B.}\ \bibnamefont
  {Nasr}}, \bibinfo {author} {\bibfnamefont {S.}~\bibnamefont {Carrasco}},
  \bibinfo {author} {\bibfnamefont {B.~E.~A.}\ \bibnamefont {Saleh}}, \bibinfo
  {author} {\bibfnamefont {A.~V.}\ \bibnamefont {Sergienko}}, \bibinfo {author}
  {\bibfnamefont {M.~C.}\ \bibnamefont {Teich}}, \bibinfo {author}
  {\bibfnamefont {J.~P.}\ \bibnamefont {Torres}}, \bibinfo {author}
  {\bibfnamefont {L.}~\bibnamefont {Torner}}, \bibinfo {author} {\bibfnamefont
  {D.~S.}\ \bibnamefont {Hum}}, \ and\ \bibinfo {author} {\bibfnamefont
  {M.~M.}\ \bibnamefont {Fejer}},\ }\href {\doibase
  10.1103/PhysRevLett.100.183601} {\bibfield  {journal} {\bibinfo  {journal}
  {Phys. Rev. Lett.}\ }\textbf {\bibinfo {volume} {100}},\ \bibinfo {pages}
  {183601} (\bibinfo {year} {2008})}\BibitemShut {NoStop}%
\bibitem [{\citenamefont {Sensarn}\ \emph {et~al.}(2010)\citenamefont
  {Sensarn}, \citenamefont {Yin},\ and\ \citenamefont {Harris}}]{Sensarn2010}%
  \BibitemOpen
  \bibfield  {author} {\bibinfo {author} {\bibfnamefont {S.}~\bibnamefont
  {Sensarn}}, \bibinfo {author} {\bibfnamefont {G.~Y.}\ \bibnamefont {Yin}}, \
  and\ \bibinfo {author} {\bibfnamefont {S.~E.}\ \bibnamefont {Harris}},\
  }\href {\doibase 10.1103/PhysRevLett.104.253602} {\bibfield  {journal}
  {\bibinfo  {journal} {Phys. Rev. Lett.}\ }\textbf {\bibinfo {volume} {104}},\
  \bibinfo {pages} {253602} (\bibinfo {year} {2010})}\BibitemShut {NoStop}%
\bibitem [{\citenamefont {Odele}\ \emph {et~al.}(2015)\citenamefont {Odele},
  \citenamefont {Lukens}, \citenamefont {Jaramillo-Villegas}, \citenamefont
  {Langrock}, \citenamefont {Fejer}, \citenamefont {Leaird},\ and\
  \citenamefont {Weiner}}]{Odele2015}%
  \BibitemOpen
  \bibfield  {author} {\bibinfo {author} {\bibfnamefont {O.~D.}\ \bibnamefont
  {Odele}}, \bibinfo {author} {\bibfnamefont {J.~M.}\ \bibnamefont {Lukens}},
  \bibinfo {author} {\bibfnamefont {J.~A.}\ \bibnamefont {Jaramillo-Villegas}},
  \bibinfo {author} {\bibfnamefont {C.}~\bibnamefont {Langrock}}, \bibinfo
  {author} {\bibfnamefont {M.~M.}\ \bibnamefont {Fejer}}, \bibinfo {author}
  {\bibfnamefont {D.~E.}\ \bibnamefont {Leaird}}, \ and\ \bibinfo {author}
  {\bibfnamefont {A.~M.}\ \bibnamefont {Weiner}},\ }\href@noop {} {\bibfield
  {journal} {\bibinfo  {journal} {Opt. Express}\ }\textbf {\bibinfo {volume}
  {23}},\ \bibinfo {pages} {21857} (\bibinfo {year} {2015})}\BibitemShut
  {NoStop}%
\bibitem [{\citenamefont {Odele}\ \emph {et~al.}(2017)\citenamefont {Odele},
  \citenamefont {Lukens}, \citenamefont {Jaramillo-Villegas}, \citenamefont
  {Imany}, \citenamefont {Langrock}, \citenamefont {Fejer}, \citenamefont
  {Leaird},\ and\ \citenamefont {Weiner}}]{Odele2017}%
  \BibitemOpen
  \bibfield  {author} {\bibinfo {author} {\bibfnamefont {O.~D.}\ \bibnamefont
  {Odele}}, \bibinfo {author} {\bibfnamefont {J.~M.}\ \bibnamefont {Lukens}},
  \bibinfo {author} {\bibfnamefont {J.~A.}\ \bibnamefont {Jaramillo-Villegas}},
  \bibinfo {author} {\bibfnamefont {P.}~\bibnamefont {Imany}}, \bibinfo
  {author} {\bibfnamefont {C.}~\bibnamefont {Langrock}}, \bibinfo {author}
  {\bibfnamefont {M.~M.}\ \bibnamefont {Fejer}}, \bibinfo {author}
  {\bibfnamefont {D.~E.}\ \bibnamefont {Leaird}}, \ and\ \bibinfo {author}
  {\bibfnamefont {A.~M.}\ \bibnamefont {Weiner}},\ }\href@noop {} {\bibfield
  {journal} {\bibinfo  {journal} {APL Photon.}\ }\textbf {\bibinfo {volume}
  {2}},\ \bibinfo {pages} {011301} (\bibinfo {year} {2017})}\BibitemShut
  {NoStop}%
\bibitem [{\citenamefont {Bra\'{n}czyk}\ \emph {et~al.}(2011)\citenamefont
  {Bra\'{n}czyk}, \citenamefont {Fedrizzi}, \citenamefont {Stace},
  \citenamefont {Ralph},\ and\ \citenamefont {White}}]{Branczyk2011}%
  \BibitemOpen
  \bibfield  {author} {\bibinfo {author} {\bibfnamefont {A.~M.}\ \bibnamefont
  {Bra\'{n}czyk}}, \bibinfo {author} {\bibfnamefont {A.}~\bibnamefont
  {Fedrizzi}}, \bibinfo {author} {\bibfnamefont {T.~M.}\ \bibnamefont {Stace}},
  \bibinfo {author} {\bibfnamefont {T.~C.}\ \bibnamefont {Ralph}}, \ and\
  \bibinfo {author} {\bibfnamefont {A.~G.}\ \bibnamefont {White}},\ }\href
  {\doibase 10.1364/OE.19.000055} {\bibfield  {journal} {\bibinfo  {journal}
  {Opt. Express}\ }\textbf {\bibinfo {volume} {19}},\ \bibinfo {pages} {55}
  (\bibinfo {year} {2011})}\BibitemShut {NoStop}%
\bibitem [{\citenamefont {Dixon}\ \emph {et~al.}(2013)\citenamefont {Dixon},
  \citenamefont {Shapiro},\ and\ \citenamefont {Wong}}]{BenDixon2013}%
  \BibitemOpen
  \bibfield  {author} {\bibinfo {author} {\bibfnamefont {P.~B.}\ \bibnamefont
  {Dixon}}, \bibinfo {author} {\bibfnamefont {J.~H.}\ \bibnamefont {Shapiro}},
  \ and\ \bibinfo {author} {\bibfnamefont {F.~N.~C.}\ \bibnamefont {Wong}},\
  }\href {\doibase 10.1364/OE.21.005879} {\bibfield  {journal} {\bibinfo
  {journal} {Opt. Express}\ }\textbf {\bibinfo {volume} {21}},\ \bibinfo
  {pages} {5879} (\bibinfo {year} {2013})}\BibitemShut {NoStop}%
\bibitem [{\citenamefont {Chen}\ \emph {et~al.}(2017)\citenamefont {Chen},
  \citenamefont {Bo}, \citenamefont {Niu}, \citenamefont {Xu}, \citenamefont
  {Zhang}, \citenamefont {Shapiro},\ and\ \citenamefont {Wong}}]{Chen2017}%
  \BibitemOpen
  \bibfield  {author} {\bibinfo {author} {\bibfnamefont {C.}~\bibnamefont
  {Chen}}, \bibinfo {author} {\bibfnamefont {C.}~\bibnamefont {Bo}}, \bibinfo
  {author} {\bibfnamefont {M.~Y.}\ \bibnamefont {Niu}}, \bibinfo {author}
  {\bibfnamefont {F.}~\bibnamefont {Xu}}, \bibinfo {author} {\bibfnamefont
  {Z.}~\bibnamefont {Zhang}}, \bibinfo {author} {\bibfnamefont {J.~H.}\
  \bibnamefont {Shapiro}}, \ and\ \bibinfo {author} {\bibfnamefont {F.~N.~C.}\
  \bibnamefont {Wong}},\ }\href {\doibase 10.1364/OE.25.007300} {\bibfield
  {journal} {\bibinfo  {journal} {Opt. Express}\ }\textbf {\bibinfo {volume}
  {25}},\ \bibinfo {pages} {7300} (\bibinfo {year} {2017})}\BibitemShut
  {NoStop}%
\bibitem [{SFW()}]{SFWM}%
  \BibitemOpen
  \href@noop {} {\bibinfo  {journal} {This analysis also holds for spontaneous
  four-wave mixing (SFWM), with the single $\alpha$ terms replaced by sums of
  products of pump amplitudes. Processes with identical total biphoton energy
  will still depend on the same pump contributions, so that SFWM will face the
  same general limits to arbitrary state production as SPDC}\ }\BibitemShut
  {NoStop}%
\bibitem [{\citenamefont {Imany}\ \emph
  {et~al.}(2018{\natexlab{b}})\citenamefont {Imany}, \citenamefont {Odele},
  \citenamefont {Alshaykh}, \citenamefont {Lu}, \citenamefont {Leaird},\ and\
  \citenamefont {Weiner}}]{imany2018frequency}%
  \BibitemOpen
\bibfield  {journal} {  }\bibfield  {author} {\bibinfo {author} {\bibfnamefont
  {P.}~\bibnamefont {Imany}}, \bibinfo {author} {\bibfnamefont {O.~D.}\
  \bibnamefont {Odele}}, \bibinfo {author} {\bibfnamefont {M.~S.}\ \bibnamefont
  {Alshaykh}}, \bibinfo {author} {\bibfnamefont {H.-H.}\ \bibnamefont {Lu}},
  \bibinfo {author} {\bibfnamefont {D.~E.}\ \bibnamefont {Leaird}}, \ and\
  \bibinfo {author} {\bibfnamefont {A.~M.}\ \bibnamefont {Weiner}},\
  }\href@noop {} {\bibfield  {journal} {\bibinfo  {journal} {Opt. Lett.}\
  }\textbf {\bibinfo {volume} {43}},\ \bibinfo {pages} {2760} (\bibinfo {year}
  {2018}{\natexlab{b}})}\BibitemShut {NoStop}%
\bibitem [{Sup()}]{Supplement}%
  \BibitemOpen
  \href@noop {} {\bibinfo  {journal} {See the Supplemental Material for
  derivation of coincidence probability in mutually unbiased basis as a
  function of temporal delay traversed by the photons and RF modulation phase}\
  }\BibitemShut {NoStop}%
\bibitem [{\citenamefont {Monmayrant}\ and\ \citenamefont
  {Chatel}(2004)}]{Monmayrant2004}%
  \BibitemOpen
\bibfield  {journal} {  }\bibfield  {author} {\bibinfo {author} {\bibfnamefont
  {A.}~\bibnamefont {Monmayrant}}\ and\ \bibinfo {author} {\bibfnamefont
  {B.}~\bibnamefont {Chatel}},\ }\href@noop {} {\bibfield  {journal} {\bibinfo
  {journal} {Rev. Sci. Instrum.}\ }\textbf {\bibinfo {volume} {75}},\ \bibinfo
  {pages} {2668} (\bibinfo {year} {2004})}\BibitemShut {NoStop}%
\bibitem [{\citenamefont {Willits}\ \emph {et~al.}(2012)\citenamefont
  {Willits}, \citenamefont {Weiner},\ and\ \citenamefont
  {Cundiff}}]{Willits2012}%
  \BibitemOpen
  \bibfield  {author} {\bibinfo {author} {\bibfnamefont {J.~T.}\ \bibnamefont
  {Willits}}, \bibinfo {author} {\bibfnamefont {A.~M.}\ \bibnamefont {Weiner}},
  \ and\ \bibinfo {author} {\bibfnamefont {S.~T.}\ \bibnamefont {Cundiff}},\
  }\href@noop {} {\bibfield  {journal} {\bibinfo  {journal} {Opt. Express}\
  }\textbf {\bibinfo {volume} {20}},\ \bibinfo {pages} {3110} (\bibinfo {year}
  {2012})}\BibitemShut {NoStop}%
\bibitem [{\citenamefont {Lukens}\ \emph {et~al.}(2020)\citenamefont {Lukens},
  \citenamefont {Law}, \citenamefont {Jasra},\ and\ \citenamefont
  {Lougovski}}]{lukens2020practical}%
  \BibitemOpen
  \bibfield  {author} {\bibinfo {author} {\bibfnamefont {J.~M.}\ \bibnamefont
  {Lukens}}, \bibinfo {author} {\bibfnamefont {K.~J.}\ \bibnamefont {Law}},
  \bibinfo {author} {\bibfnamefont {A.}~\bibnamefont {Jasra}}, \ and\ \bibinfo
  {author} {\bibfnamefont {P.}~\bibnamefont {Lougovski}},\ }\href@noop {}
  {\bibfield  {journal} {\bibinfo  {journal} {New J. Phys.}\ }\textbf {\bibinfo
  {volume} {22}},\ \bibinfo {pages} {063038} (\bibinfo {year}
  {2020})}\BibitemShut {NoStop}%
\bibitem [{\citenamefont {Lu}\ \emph {et~al.}(2021)\citenamefont {Lu},
  \citenamefont {Myilswamy}, \citenamefont {Bennink}, \citenamefont {Seshadri},
  \citenamefont {Alshaykh}, \citenamefont {Liu}, \citenamefont {Kippenberg},
  \citenamefont {Leaird}, \citenamefont {Weiner},\ and\ \citenamefont
  {Lukens}}]{lu2021full}%
  \BibitemOpen
  \bibfield  {author} {\bibinfo {author} {\bibfnamefont {H.-H.}\ \bibnamefont
  {Lu}}, \bibinfo {author} {\bibfnamefont {K.~V.}\ \bibnamefont {Myilswamy}},
  \bibinfo {author} {\bibfnamefont {R.~S.}\ \bibnamefont {Bennink}}, \bibinfo
  {author} {\bibfnamefont {S.}~\bibnamefont {Seshadri}}, \bibinfo {author}
  {\bibfnamefont {M.~S.}\ \bibnamefont {Alshaykh}}, \bibinfo {author}
  {\bibfnamefont {J.}~\bibnamefont {Liu}}, \bibinfo {author} {\bibfnamefont
  {T.~J.}\ \bibnamefont {Kippenberg}}, \bibinfo {author} {\bibfnamefont
  {D.~E.}\ \bibnamefont {Leaird}}, \bibinfo {author} {\bibfnamefont {A.~M.}\
  \bibnamefont {Weiner}}, \ and\ \bibinfo {author} {\bibfnamefont {J.~M.}\
  \bibnamefont {Lukens}},\ }\href@noop {} {\bibfield  {journal} {\bibinfo
  {journal} {arXiv:2108.04124}\ } (\bibinfo {year} {2021})}\BibitemShut
  {NoStop}%
\bibitem [{\citenamefont {Lu}\ \emph {et~al.}(2022)\citenamefont {Lu},
  \citenamefont {Lingaraju}, \citenamefont {Leaird}, \citenamefont {Weiner},\
  and\ \citenamefont {Lukens}}]{lu2022high}%
  \BibitemOpen
  \bibfield  {author} {\bibinfo {author} {\bibfnamefont {H.-H.}\ \bibnamefont
  {Lu}}, \bibinfo {author} {\bibfnamefont {N.~B.}\ \bibnamefont {Lingaraju}},
  \bibinfo {author} {\bibfnamefont {D.~E.}\ \bibnamefont {Leaird}}, \bibinfo
  {author} {\bibfnamefont {A.~M.}\ \bibnamefont {Weiner}}, \ and\ \bibinfo
  {author} {\bibfnamefont {J.~M.}\ \bibnamefont {Lukens}},\ }\href {\doibase
  10.1364/OE.454677} {\bibfield  {journal} {\bibinfo  {journal} {Opt. Express}\
  }\textbf {\bibinfo {volume} {30}},\ \bibinfo {pages} {10126} (\bibinfo {year}
  {2022})}\BibitemShut {NoStop}%
\bibitem [{\citenamefont {Giovannetti}\ \emph {et~al.}(2004)\citenamefont
  {Giovannetti}, \citenamefont {Lloyd},\ and\ \citenamefont
  {Maccone}}]{giovannetti2004quantum}%
  \BibitemOpen
  \bibfield  {author} {\bibinfo {author} {\bibfnamefont {V.}~\bibnamefont
  {Giovannetti}}, \bibinfo {author} {\bibfnamefont {S.}~\bibnamefont {Lloyd}},
  \ and\ \bibinfo {author} {\bibfnamefont {L.}~\bibnamefont {Maccone}},\
  }\href@noop {} {\bibfield  {journal} {\bibinfo  {journal} {Science}\ }\textbf
  {\bibinfo {volume} {306}},\ \bibinfo {pages} {1330} (\bibinfo {year}
  {2004})}\BibitemShut {NoStop}%
\bibitem [{\citenamefont {Quan}\ \emph {et~al.}(2020)\citenamefont {Quan},
  \citenamefont {Dong}, \citenamefont {Xiang}, \citenamefont {Li},
  \citenamefont {Liu},\ and\ \citenamefont {Zhang}}]{nonlocalquan2020high}%
  \BibitemOpen
  \bibfield  {author} {\bibinfo {author} {\bibfnamefont {R.}~\bibnamefont
  {Quan}}, \bibinfo {author} {\bibfnamefont {R.}~\bibnamefont {Dong}}, \bibinfo
  {author} {\bibfnamefont {X.}~\bibnamefont {Xiang}}, \bibinfo {author}
  {\bibfnamefont {B.}~\bibnamefont {Li}}, \bibinfo {author} {\bibfnamefont
  {T.}~\bibnamefont {Liu}}, \ and\ \bibinfo {author} {\bibfnamefont
  {S.}~\bibnamefont {Zhang}},\ }\href@noop {} {\bibfield  {journal} {\bibinfo
  {journal} {Rev. Sci. Instrum.}\ }\textbf {\bibinfo {volume} {91}},\ \bibinfo
  {pages} {123109} (\bibinfo {year} {2020})}\BibitemShut {NoStop}%
\bibitem [{\citenamefont {Seshadri}\ \emph {et~al.}(2022)\citenamefont
  {Seshadri}, \citenamefont {Lingaraju}, \citenamefont {Lu}, \citenamefont
  {Imany}, \citenamefont {Leaird},\ and\ \citenamefont
  {Weiner}}]{seshadri2022nonlocal}%
  \BibitemOpen
  \bibfield  {author} {\bibinfo {author} {\bibfnamefont {S.}~\bibnamefont
  {Seshadri}}, \bibinfo {author} {\bibfnamefont {N.}~\bibnamefont {Lingaraju}},
  \bibinfo {author} {\bibfnamefont {H.-H.}\ \bibnamefont {Lu}}, \bibinfo
  {author} {\bibfnamefont {P.}~\bibnamefont {Imany}}, \bibinfo {author}
  {\bibfnamefont {D.~E.}\ \bibnamefont {Leaird}}, \ and\ \bibinfo {author}
  {\bibfnamefont {A.~M.}\ \bibnamefont {Weiner}},\ }\href@noop {} {\bibfield
  {journal} {\bibinfo  {journal} {arXiv:2202.11816}\ } (\bibinfo {year}
  {2022})}\BibitemShut {NoStop}%
\bibitem [{\citenamefont {Kuzucu}\ \emph {et~al.}(2005)\citenamefont {Kuzucu},
  \citenamefont {Fiorentino}, \citenamefont {Albota}, \citenamefont {Wong},\
  and\ \citenamefont {K{\"a}rtner}}]{kuzucu2005two}%
  \BibitemOpen
  \bibfield  {author} {\bibinfo {author} {\bibfnamefont {O.}~\bibnamefont
  {Kuzucu}}, \bibinfo {author} {\bibfnamefont {M.}~\bibnamefont {Fiorentino}},
  \bibinfo {author} {\bibfnamefont {M.~A.}\ \bibnamefont {Albota}}, \bibinfo
  {author} {\bibfnamefont {F.~N.}\ \bibnamefont {Wong}}, \ and\ \bibinfo
  {author} {\bibfnamefont {F.~X.}\ \bibnamefont {K{\"a}rtner}},\ }\href@noop {}
  {\bibfield  {journal} {\bibinfo  {journal} {Phys. Rev. Lett.}\ }\textbf
  {\bibinfo {volume} {94}},\ \bibinfo {pages} {083601} (\bibinfo {year}
  {2005})}\BibitemShut {NoStop}%
\bibitem [{\citenamefont {Kuzucu}\ \emph {et~al.}(2008)\citenamefont {Kuzucu},
  \citenamefont {Wong}, \citenamefont {Kurimura},\ and\ \citenamefont
  {Tovstonog}}]{kuzucu2008joint}%
  \BibitemOpen
  \bibfield  {author} {\bibinfo {author} {\bibfnamefont {O.}~\bibnamefont
  {Kuzucu}}, \bibinfo {author} {\bibfnamefont {F.~N.}\ \bibnamefont {Wong}},
  \bibinfo {author} {\bibfnamefont {S.}~\bibnamefont {Kurimura}}, \ and\
  \bibinfo {author} {\bibfnamefont {S.}~\bibnamefont {Tovstonog}},\ }\href@noop
  {} {\bibfield  {journal} {\bibinfo  {journal} {Phys. Rev. Lett.}\ }\textbf
  {\bibinfo {volume} {101}},\ \bibinfo {pages} {153602} (\bibinfo {year}
  {2008})}\BibitemShut {NoStop}%
\bibitem [{\citenamefont {Zhao}\ \emph {et~al.}(2021)\citenamefont {Zhao},
  \citenamefont {Zhang}, \citenamefont {Liu}, \citenamefont {Guan},
  \citenamefont {Zhang}, \citenamefont {Li}, \citenamefont {Bai}, \citenamefont
  {Li}, \citenamefont {Liu}, \citenamefont {You}, \citenamefont {Zhang},
  \citenamefont {Fan}, \citenamefont {Xu}, \citenamefont {Zhang},\ and\
  \citenamefont {Pan}}]{zhao2021field}%
  \BibitemOpen
  \bibfield  {author} {\bibinfo {author} {\bibfnamefont {S.-R.}\ \bibnamefont
  {Zhao}}, \bibinfo {author} {\bibfnamefont {Y.-Z.}\ \bibnamefont {Zhang}},
  \bibinfo {author} {\bibfnamefont {W.-Z.}\ \bibnamefont {Liu}}, \bibinfo
  {author} {\bibfnamefont {J.-Y.}\ \bibnamefont {Guan}}, \bibinfo {author}
  {\bibfnamefont {W.}~\bibnamefont {Zhang}}, \bibinfo {author} {\bibfnamefont
  {C.-L.}\ \bibnamefont {Li}}, \bibinfo {author} {\bibfnamefont
  {B.}~\bibnamefont {Bai}}, \bibinfo {author} {\bibfnamefont {M.-H.}\
  \bibnamefont {Li}}, \bibinfo {author} {\bibfnamefont {Y.}~\bibnamefont
  {Liu}}, \bibinfo {author} {\bibfnamefont {L.}~\bibnamefont {You}}, \bibinfo
  {author} {\bibfnamefont {J.}~\bibnamefont {Zhang}}, \bibinfo {author}
  {\bibfnamefont {J.}~\bibnamefont {Fan}}, \bibinfo {author} {\bibfnamefont
  {F.}~\bibnamefont {Xu}}, \bibinfo {author} {\bibfnamefont {Q.}~\bibnamefont
  {Zhang}}, \ and\ \bibinfo {author} {\bibfnamefont {J.-W.}\ \bibnamefont
  {Pan}},\ }\href {\doibase 10.1103/PhysRevX.11.031009} {\bibfield  {journal}
  {\bibinfo  {journal} {Phys. Rev. X}\ }\textbf {\bibinfo {volume} {11}},\
  \bibinfo {pages} {031009} (\bibinfo {year} {2021})}\BibitemShut {NoStop}%
\bibitem [{\citenamefont {Giovannetti}\ \emph {et~al.}(2011)\citenamefont
  {Giovannetti}, \citenamefont {Lloyd},\ and\ \citenamefont
  {Maccone}}]{giovannetti2011advances}%
  \BibitemOpen
  \bibfield  {author} {\bibinfo {author} {\bibfnamefont {V.}~\bibnamefont
  {Giovannetti}}, \bibinfo {author} {\bibfnamefont {S.}~\bibnamefont {Lloyd}},
  \ and\ \bibinfo {author} {\bibfnamefont {L.}~\bibnamefont {Maccone}},\
  }\href@noop {} {\bibfield  {journal} {\bibinfo  {journal} {Nat. Photon.}\
  }\textbf {\bibinfo {volume} {5}},\ \bibinfo {pages} {222} (\bibinfo {year}
  {2011})}\BibitemShut {NoStop}%
\bibitem [{\citenamefont {Weiner}(2009)}]{weiner2011ultrafast}%
  \BibitemOpen
  \bibfield  {author} {\bibinfo {author} {\bibfnamefont {A.~M.}\ \bibnamefont
  {Weiner}},\ }\href@noop {} {\emph {\bibinfo {title} {Ultrafast Optics}}}\
  (\bibinfo  {publisher} {Wiley},\ \bibinfo {year} {2009})\BibitemShut
  {NoStop}%
\bibitem [{\citenamefont {Pe'er}\ \emph {et~al.}(2005)\citenamefont {Pe'er},
  \citenamefont {Dayan}, \citenamefont {Friesem},\ and\ \citenamefont
  {Silberberg}}]{pe2005temporal}%
  \BibitemOpen
  \bibfield  {author} {\bibinfo {author} {\bibfnamefont {A.}~\bibnamefont
  {Pe'er}}, \bibinfo {author} {\bibfnamefont {B.}~\bibnamefont {Dayan}},
  \bibinfo {author} {\bibfnamefont {A.~A.}\ \bibnamefont {Friesem}}, \ and\
  \bibinfo {author} {\bibfnamefont {Y.}~\bibnamefont {Silberberg}},\
  }\href@noop {} {\bibfield  {journal} {\bibinfo  {journal} {Phys. Rev. Lett.}\
  }\textbf {\bibinfo {volume} {94}},\ \bibinfo {pages} {073601} (\bibinfo
  {year} {2005})}\BibitemShut {NoStop}%
\bibitem [{\citenamefont {Chun-Yan}\ \emph {et~al.}(2005)\citenamefont
  {Chun-Yan}, \citenamefont {Hong-Yu}, \citenamefont {Yan},\ and\ \citenamefont
  {Fu-Guo}}]{chun2005secure}%
  \BibitemOpen
  \bibfield  {author} {\bibinfo {author} {\bibfnamefont {L.}~\bibnamefont
  {Chun-Yan}}, \bibinfo {author} {\bibfnamefont {Z.}~\bibnamefont {Hong-Yu}},
  \bibinfo {author} {\bibfnamefont {W.}~\bibnamefont {Yan}}, \ and\ \bibinfo
  {author} {\bibfnamefont {D.}~\bibnamefont {Fu-Guo}},\ }\href@noop {}
  {\bibfield  {journal} {\bibinfo  {journal} {Chin. Phys. Lett.}\ }\textbf
  {\bibinfo {volume} {22}},\ \bibinfo {pages} {1049} (\bibinfo {year}
  {2005})}\BibitemShut {NoStop}%
\bibitem [{\citenamefont {Liscidini}\ and\ \citenamefont
  {Sipe}(2019)}]{liscidini2019scalable}%
  \BibitemOpen
  \bibfield  {author} {\bibinfo {author} {\bibfnamefont {M.}~\bibnamefont
  {Liscidini}}\ and\ \bibinfo {author} {\bibfnamefont {J.}~\bibnamefont
  {Sipe}},\ }\href@noop {} {\bibfield  {journal} {\bibinfo  {journal} {Opt.
  Lett.}\ }\textbf {\bibinfo {volume} {44}},\ \bibinfo {pages} {2625} (\bibinfo
  {year} {2019})}\BibitemShut {NoStop}%
\end{thebibliography}%
\vspace{2mm}

\end{document}